# Nonlinear dynamics and fluctuations in biological systems

## Kumulative Habilitationsschrift

im Fach
Theoretische Physik

vorgelegt
der Fakultät Mathematik und Naturwissenschaften
der Technischen Universität Dresden
von

Dr. rer. nat. Benjamin Friedrich
geboren am 22.07.1979 in Rostock

eingereicht am 30.05.2016;
Wissenschaftlicher Vortrag und Aussprache am 11.12.2017

Gutachter:
Prof. Dr. Eberhard Bodenschatz
Prof. Dr. Erwin Frey
Prof. Dr. Stephan Grill

Technische Universität Dresden

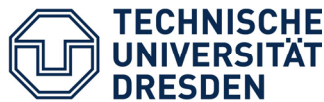

Die Habilitationsschrift wurde in der Zeit von Oktober 2011 bis Mai 2016
am Max-Planck-Institut für Physik komplexer Systeme in Dresden angefertigt.

# Table of Contents

# Abstract

The present habilitation thesis in theoretical biological physics addresses two central dynamical processes in cells and organisms: (i) active motility and motility control and (ii) self-organized pattern formation. The unifying theme is the nonlinear dynamics of biological function and its robustness in the presence of strong fluctuations, structural variations, and external perturbations.

We theoretically investigate motility control at the cellular scale, using cilia and flagella as ideal model system. Cilia and flagella are highly conserved slender cell appendages that exhibit spontaneous bending waves. This flagellar beat represents a prime example of a chemo-mechanical oscillator, which is driven by the collective dynamics of molecular motors inside the flagellar axoneme. We study the nonlinear dynamics of flagellar swimming, steering, and synchronization, which encompasses shape control of the flagellar beat by chemical signals and mechanical forces. Mechanical forces can synchronize collections of flagella to beat at a common frequency, despite active motor noise that tends to randomize flagellar synchrony. In Chapter 2, we present a new physical mechanism for flagellar synchronization by mechanical self-stabilization that applies to free-swimming flagellated cells. This new mechanism is independent of direct hydrodynamic interactions between flagella. Comparison with experimental data provided by experimental collaboration partners in the laboratory of J. Howard (Yale, New Haven) confirmed our new mechanism in the model organism of the unicellular green alga *Chlamydomonas*. Further, we characterize the beating flagellum as a noisy oscillator. Using a minimal model of collective motor dynamics, we argue that measured non-equilibrium fluctuations of the flagellar beat result from stochastic motor dynamics at the molecular scale. Noise and mechanical coupling are antagonists for flagellar synchronization.

In addition to the control of the flagellar beat by mechanical forces, we study the control of the flagellar beat by chemical signals in the context of sperm chemotaxis. We characterize a fundamental paradigm for navigation in external concentration gradients that relies on active swimming along helical paths. In this helical chemotaxis, the direction of a spatial concentration gradient becomes encoded in the phase of an oscillatory chemical signal. Helical chemotaxis represents a distinct gradient-sensing strategy, which is different from bacterial chemotaxis. Helical chemotaxis is employed, for example, by sperm cells from marine invertebrates with external fertilization. We present a theory of sensorimotor control, which combines hydrodynamic simulations of chiral flagellar swimming with a dynamic regulation of flagellar beat shape in response to chemical signals perceived by the cell. Our theory is compared to three-dimensional tracking experiments of sperm chemotaxis performed by the laboratory of U. B. Kaupp (CAESAR, Bonn).

In addition to motility control, we investigate in Chapter 3 self-organized pattern formation in two selected biological systems at the cell and organism scale, respectively. On the cellular scale, we present a minimal physical mechanism for the spontaneous self-assembly of periodic cytoskeletal patterns, as observed in myofibrils in striated muscle cells. This minimal mechanism relies on the interplay of a passive coarsening process of crosslinked actin clusters and active cytoskeletal forces. This mechanism of cytoskeletal pattern formation exemplifies how local interactions can generate large-scale spatial order in active systems.

On the organism scale, we present an extension of Turing's framework for self-organized pattern formation that is capable of a proportionate scaling of steady-state patterns with system size. This new mechanism does not require any pre-patterning clues and can restore proportional patterns in regeneration scenarios. We analytically derive the hierarchy of steady-state patterns and analyze their stability and basins of attraction. We demonstrate that this scaling mechanism is structurally robust. Applications to the growth and regeneration dynamics in flatworms are discussed (experiments by J. Rink, MPI CBG, Dresden).

# Zusammenfassung [Abstract in German]

Das Thema der vorliegenden Habilitationsschrift in Theoretischer Biologischer Physik ist die nichtlineare Dynamik funktionaler biologischer Systeme und deren Robustheit gegenüber Fluktuationen und äußeren Störungen. Wir entwickeln hierzu theoretische Beschreibungen für zwei grundlegende biologische Prozesse: (i) die zell-autonome Kontrolle aktiver Bewegung, sowie (ii) selbstorganisierte Musterbildung in Zellen und Organismen.

In Kapitel 2, untersuchen wir Bewegungskontrolle auf zellulärer Ebene am Modelsystem von Zilien und Geißeln. Spontane Biegewellen dieser dünnen Zellfortsätze ermöglichen es eukaryotischen Zellen, in einer Flüssigkeit zu schwimmen. Wir beschreiben einen neuen physikalischen Mechanismus für die Synchronisation zweier schlagender Geißeln, unabhängig von direkten hydrodynamischen Wechselwirkungen. Der Vergleich mit experimentellen Daten, zur Verfügung gestellt von unseren experimentellen Kooperationspartnern im Labor von J. Howard (Yale, New Haven), bestätigt diesen neuen Mechanismus im Modellorganismus der einzelligen Grünalge *Chlamydomonas*. Der Gegenspieler dieser Synchronisation durch mechanische Kopplung sind Fluktuationen. Wir bestimmen erstmals Nichtgleichgewichts-Fluktuationen des Geißel-Schlags direkt, wofür wir eine neue Analyse-Methode der Grenzzykel-Rekonstruktion entwickeln. Die von uns gemessenen Fluktuationen entstehen mutmaßlich durch die stochastische Dynamik molekularen Motoren im Innern der Geißeln, welche auch den Geißelschlag antreiben. Um die statistische Physik dieser Nichtgleichgewichts-Fluktuationen zu verstehen, entwickeln wir eine analytische Theorie der Fluktuationen in einem minimalen Modell kollektiver Motor-Dynamik. Zusätzlich zur Regulation des Geißelschlags durch mechanische Kräfte untersuchen wir dessen Regulation durch chemische Signale am Modell der Chemotaxis von Spermien-Zellen. Dabei charakterisieren wir einen grundlegenden Mechanismus für die Navigation in externen Konzentrationsgradienten. Dieser Mechanismus beruht auf dem aktiven Schwimmen entlang von Spiralbahnen, wodurch ein räumlicher Konzentrationsgradient in der Phase eines oszillierenden chemischen Signals kodiert wird. Dieser Chemotaxis-Mechanismus unterscheidet sich grundlegend vom bekannten Chemotaxis-Mechanismus von Bakterien. Wir entwickeln eine Theorie der senso-motorischen Steuerung des Geißelschlags während der Spermien-Chemotaxis. Vorhersagen dieser Theorie werden durch Experimente der Gruppe von U.B. Kaupp (CAESAR, Bonn) quantitativ bestätigt.

In Kapitel 3, untersuchen wir selbstorganisierte Strukturbildung in zwei ausgewählten biologischen Systemen. Auf zellulärer Ebene schlagen wir einen einfachen physikalischen Mechanismus vor für die spontane Selbstorganisation von periodischen Zellskelett-Strukturen, wie sie sich z.B. in den Myofibrillen gestreifter Muskelzellen finden. Dieser Mechanismus zeigt exemplarisch auf, wie allein durch lokale Wechselwirkungen räumliche Ordnung auf größeren Längenskalen in einem Nichtgleichgewichtssystem entstehen kann. Auf der Ebene des Organismus stellen wir eine Erweiterung der Turingschen Theorie für selbstorganisierte Musterbildung vor. Wir beschreiben eine neue Klasse von Musterbildungssystemen, welche selbst-organisierte Muster erzeugt, die mit der Systemgröße skalieren. Dieser neue Mechanismus erfordert weder eine vorgegebene Kompartimentalisierung des Systems noch spezielle Randbedingungen. Insbesondere kann dieser Mechanismus proportionale Muster wiederherstellen, wenn Teile des Systems amputiert werden. Wir bestimmen analytisch die Hierarchie aller stationären Muster und analysieren deren Stabilität und Einzugsgebiete. Damit können wir zeigen, dass dieser Skalierungs-Mechanismus strukturell robust ist bezüglich Variationen von Parametern und sogar funktionalen Beziehungen zwischen dynamischen Variablen. Zusammen mit Kollaborationspartnern im Labor von J. Rink (MPI CBG, Dresden) diskutieren wir Anwendungen auf das Wachstum von Plattwürmern und deren Regeneration in Amputations-Experimenten.

# Acknowledgement

Foremost, I thank Stephan Grill for being the mentor for my habilitation. He made it possible that I could offer lectures on hydrodynamics, elasticity theory and nonlinear dynamics at the Technical University of Dresden in the years 2014-2016. Furthermore, I wholeheartedly thank Frank Jülicher for his continuous support and many inspiring scientific discussions.

I am strongly indebted to a large number of experimental collaboration partners, who generously shared primary data as well as their expert knowledge of biological model systems. Specifically, for the publications selected in this thesis, I had the pleasure to work with Luis Alvarez, Veikko F. Geyer, Jonathon Howard, U. Benjamin Kaupp, Jochen C. Rink, and Tom Stückemann. In several occasions, experimental collaboration partners performed experiments that were proposed by me and directly aimed to test theoretical ideas, for which I am most grateful.

I had and still have the privilege to work with three great PhD students, Steffen Werner, Gary S. Klindt, and André Scholich. Most of our joint results are not yet included in this thesis. I also thank all colleagues and friends at the Max Planck Institute for the Physics of Complex Systems, as well as at the Max Planck Institute for Molecular Cell Biology and Genetics in Dresden, for generating a thriving research environment. A special 'thank you' goes to Johannes Baumgart, Veikko Geyer, and Rabea Seyboldt for their critical reading of this thesis. Marius Asal helped formatting the bibliography.

I thank my wife Elisabeth Fischer-Friedrich, herself a scientist, for her constructive feedback and practical advice over many years. In fact, she is co-author of one publication in this thesis.

Finally, I gratefully acknowledge funding from the Max Planck Society and the German Science Foundation.

# 1 Introduction

## 1.1 Overview of the thesis

In this habilitation thesis, we present systems-level theoretical descriptions of motility control and self-organized pattern formation in cells and tissues. The overarching theme is the nonlinear dynamics of biological function and its robustness in the presence of strong fluctuations, structural variations, and external perturbations. As examples of biological function, we focus on two central dynamical processes in cells and organisms: (i) active motility and motility control of motile cells and (ii) self-organized pattern formation in cells and tissues. Cell motility and motility control is studied in the model system of cilia and flagella, a highly conserved motile cell appendage of eukaryotic cells[1–3], which is regulated by chemical and mechanical cues. This flagellar control facilitates flagellar synchronization[4-6] and navigation of flagellated swimmers[7–9]. The second theme, pattern formation, is studied both at the sub-cellular scale in the context of self-organization of cytoskeletal filaments into regular myofibrillar patterns[10], and at the organism scale. There, we present a new mechanism for the dynamic scaling of self-organized Turing patterns[11].

**Cell motility and motility control.** We study the nonlinear dynamics of the eukaryotic flagellum, a slender cell appendage capable of spontaneous bending waves, which propels cellular microswimmers and pumps fluids in the human body. The rhythmic beat of eukaryotic flagella represents a prime example of a chemo-mechanical biological oscillator. In publication 2.1, we study the emergent dynamics that arises from the interactions between several flagella. We identified a novel mechanism of synchronization in pairs of beating flagella, which applies to free-swimming, bi-flagellated cells. This synchronization mechanism relies on a closed feedback loop between flagellar dynamics and self-motion of the cell. This novel synchronization mechanism of mechanical self-stabilization is different from a previous mechanism that had been widely discussed in the field. This alternative mechanism proposed direct hydrodynamic interactions between flagella as the primary cause of flagellar synchronization. We show that both mechanisms can synchronize pairs of flagella, yet the new mechanism is predicted to dominate in free-swimming cells.

In publication 2.2, we extend the conceptual theoretical description of flagellar synchronization developed in the previous publication towards a full quantitative description of flagellar swimming and synchronization in the model system of a swimming unicellular alga, *Chlamydomonas*. For that aim, we combine a minimal description of flagellar beat dynamics that coarse-grains active processes inside the flagellum in terms of an active driving force with a full hydrodynamic treatment of cellular swimming. We present a one-to-one comparison between theoretical results and experimental measurements that have been conducted by our collaboration partners in the Howard laboratory (now Yale University, New Haven). Using our quantitative description, we were able to quantitatively predict the swimming dynamics as well as a force-velocity relation of flagellar oscillations. Of note, the description is free from adjustable parameters: it had been fully parameterized by one set of experimental data (of synchronized beating), allowing us to make quantitative predictions that we could test against a second, complementary set of data (of desynchronized beating). The comparison of theory and experiment validates the theory and highlights the predictive power of theory, which in this case preceded the experiments.

In publication 2.3, we characterize the beating flagellum as a noisy oscillator. We present direct measurements of flagellar phase and amplitude fluctuations. These fluctuations are of active nature and surpass the contribution from thermal noise by orders of magnitude. Active fluctuations are a hallmark of active dynamics far from thermal equilibrium. In addition to this analysis of experimental data, which was guided by theoretical concepts of a noisy Hopf oscillator, we provide a theory of noisy motor oscillations. Thereby, we are able to explain the observed active fluctuations as the result of small-number fluctuations in the activity of molecular motors. Specifically, we study a minimal model of collective dynamics of molecular motors, which gives rise to spontaneous oscillations by a dynamic instability, similar to the rhythmic flagellar beat. We demonstrate how small-number-fluctuations arising from the stochastic dynamics of individual molecular motors result in active fluctuations of collective motor oscillations, similar to those of the flagellar beat. Hence, using the flagellar beat as a model system, we demonstrate that stochastic dynamics at the molecular scale can yield to measurable implications for mesoscopic dynamics at the cellular scale. We show that flagellar amplitude fluctuations introduce stochasticity in the swimming paths of flagellated swimmers such as sperm cells. Phase fluctuations disturb flagellar synchronization, which implies a competition between active fluctuations and any mechanical coupling that tends to stabilize flagellar synchrony.

In publication 2.4, we address the control of flagellar motility by chemical signals. We characterize a chemotaxis strategy along helical paths, which is employed by sperm cells to find the egg, *e.g.* in marine invertebrates with external fertilization. There, sperm cells are able to sense signaling molecules released by the egg and to steer their swimming paths upwards a concentration gradient of these molecules. We previously postulated a generic mechanism for helical chemotaxis that relies on a closed feedback loop of sensorimotor control, linking temporal chemical signals and flagellar steering responses. This helical chemotaxis represents a distinct gradient-sensing strategy that is different from the well-studied chemotaxis of bacteria along biased random walks. Recently, a close theory-experiment collaboration with the experimental laboratory of Prof. Kaupp (CAESAR, Bonn, Germany), allowed the validation of our theory on a quantitative level. In publication 2.4, we present the results of this theory-experiment collaboration, including a comprehensive theoretical description of flagellar swimming and steering. In particular, our theory accounts for the hydrodynamics of flagellar swimming for a flagellar beat whose shape is dynamically regulated by a cellular signaling cascade. This theory, which encompasses only a small set of dynamic rules, can quantitatively account for apparently complex steering behaviors of sperm cells as observed in experiments. This includes dynamic decision making of sperm cells between two distinct steering modes in a situation-specific manner.

**Self-organized pattern formation in cells and tissues.** In addition to motility control, we study pattern formation at the cell and organism level as a second example of nonlinear dynamics in biological systems.

In publication 3.1, we address the self-assembly process of a complex motor-filament system, the myofibril, which is the key force generator in striated and cardiac muscle cells. We present a minimal mechanism by which actin filaments and bipolar myosin filaments inside a one-dimensional bundle self-organize into periodic spatial patterns, similar to those found in myofibrils. This minimal mechanism demonstrates that local interactions between micrometer-sized 'active building blocks' are capable of generating spatial order on large scales. We discuss how the polydispersity of filament lengths and the stochasticity of kinetic interactions impacts on the regularity of the emergent periodic patterns.

In publication 3.2, we study pattern formation at the organism scale. We account for a remarkable biological phenomenon, the spontaneous emergence of self-organized patterns that scale with organism size. We present a minimal model for perfect pattern scaling of a head-tail gradient in the absence of pre-patterning cues. This minimal model comprises three interacting chemical species subject to a reaction-diffusion dynamics. We analytically derive a hierarchy of self-organized and self-scaling patterns. We analyze the stability of steady-state patterns, their basin of attraction, and relaxation dynamics. For this, we apply the theory of dynamical systems to a pattern formation problem. Our theory provides a conceptual framework for pattern scaling and regeneration as observed *e.g.* in flatworms. Flatworms exhibit astonishing capabilities of reversible growth and regeneration, which are studied by our experimental collaboration partners in the Rink laboratory (MPI CBG, Dresden, Germany). Our minimal theory highlights a generic mechanism that predicts signatures of self-organized pattern scaling that can be tested in experiments conducted by our collaboration partners.

These selected publications exemplify our approach of complexity reduction in complex biological systems and the quantitative comparison of theory and experiment. In all publications, we use theoretical physics to understand the nonlinear dynamics of biological function and its robustness in the presence of non-equilibrium fluctuations.

## 1.2 What is biological physics?

While biology is traditionally concerned with the study of life, including the structure, development, and behavior of living organisms and their molecular underpinnings, physics studies fundamental interactions of energy and matter, and their motion in space and time. The subject of biological physics, living matter, constitutes a common intersection between these two natural sciences. Living matter displays novel physical phenomena with unconventional features, which are not commonly recognized in equilibrium systems. These include active motility[12], non-equilibrium fluctuations[13,14], adaptive dynamics[15–17], and self-organized pattern formation[18,19]. Biological physics studies the physical principles that underlie these phenomena. On a methodological side, biological physics comprises tools from different fields of physics: dynamical systems theory, statistical physics, and computational physics, see Figure 1.

Biological systems represent complex dynamical systems, where local interactions give rise to emergent dynamics on the system's level[20]. As a prominent example, inside cells, interacting cytoskeletal filaments self-assemble into regular structures, such as stress fibers or myofibrils characterized by nematic and smectic order[21,22]. Ensembles of molecular motor proteins exhibit collective dynamics, which drives active cell motility. On the scale of tissues, chemical and mechanical communication between cells orchestrates tissue development and homeostasis. In these examples, system-scale dynamics arises from local interactions. The description of this emergent dynamics is the realm of statistical physics, yet three practical differences between the statistical physics of living and non-living condensed matter should be noted.

- First, living systems are by definition out of equilibrium[20]. Even the maintenance of a steady-state is characterized by a continuous flux of energy and mass. Dynamics far from equilibrium implies that active fluctuations can surpass thermal fluctuations[13,14].

- Second, the number of interacting constituents in living matter are often in the range of $N = 10^2 - 10^6$, not $N \approx 6 \times 10^{23}$ as in a mole of ideal gas. Examples of such constituents include individual cytoskeletal filaments inside an animal cell that form its cytoskeleton. The comparatively small number of interacting constituents implies that small-number fluctuations proportional to $N^{1/2}$ give rise to substantial deviations from mean field dynamics.

- Third, there is not a single division line between what defines the small scale and the large scale in a biological system, see Figure 1. Rather, there is a hierarchy of coarse-graining levels: from molecules to subcellular processes to cellular dynamics to tissues to organisms and even ecological systems, see Figure 1. It is the challenge of biological physics to develop appropriate effective theoretical descriptions for a specific coarse-graining level, which can bridge from one level to the next higher level[23].

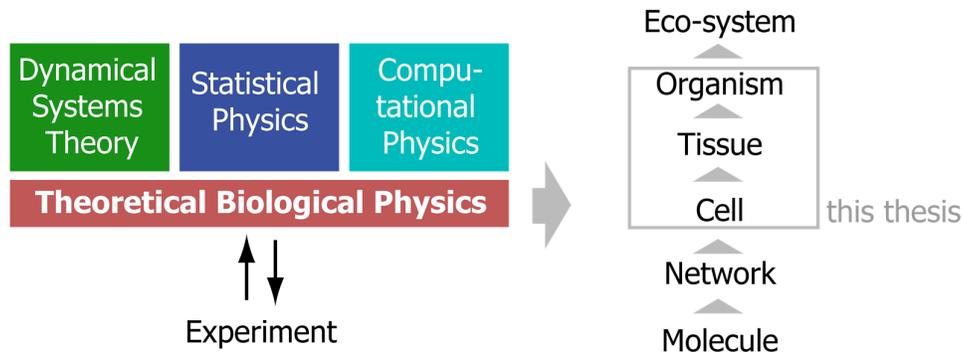

**Figure 1.** *The toolbox of theoretical biological physics.* Theoretical biological physics draws from different fields of physics. First, dynamical systems theory is indispensable to analyze effective theories of biological dynamics. Such effective theories coarse-grain dynamic processes at smaller scales and usually comprise effective degrees of freedom, *e.g.* system-level activity states. Second, statistical physics provides the framework to derive such effective theories of mesoscopic dynamics from interactions at the micro-scale. Third, computational physics enables the analysis of theoretical descriptions of biological processes at different levels of detail and complexity, which are not amenable to analytical treatment anymore. Computational methods are further needed to analyze experimental data and to quantitatively compare theory and experiment. This toolbox of theoretical biological physics is applied to identify physical mechanisms of biological function at different coarse-graining levels and length scales, ranging from subcellular dynamics up to the interactions between organisms. In this thesis, we focus on the intermediate scale of cells and organisms.

In this thesis, we employ an approach of minimality that seeks to identify those degrees of freedom of a biological system, which are absolutely needed to understand the physical principle behind a specific phenomenon, thus following the principle of Occam's razor. It is understood that any theoretical description represents an idealization of nature's complexity. We nonetheless strive for a quantitative comparison of theory and experiment. In fact, it is often only through the use of a theoretical description with a minimal number of adjustable parameters that the successful determination of parameters becomes feasible. The same applies to the falsifiability of a proposed physical mechanism of biological function. A bottom-up molecular description of biological processes is often not feasible due to the complexity of the system under study, as well as a result of our limited knowledge of its components and interactions. Even if this quantitative information were available, it is often desirable to complement bottom-up approaches by coarse-grained effective descriptions that highlight generic principles, one at a time, as studied in this thesis. We consider the crucial determinants of a theoretical description of biological system to be this choice of coarse-graining level and the choice of effective degrees of freedom, together with falsifiable assumptions on their dynamic relationships. In contrast, the actual mathematical framework used to formulate the theoretical description can often be chosen by practical considerations. Common choices include ordinary and partial differential equations, stochastic differential equations, finite difference equations, agent-based simulations, Markov models, Boolean networks, and cellular automata. Thus, the same physical idea may be cast into different specific mathematical formulations, which can often be considered equivalent on a conceptual level. The theoretical descriptions in this thesis employ stochastic nonlinear differential equations, both ordinary and partial, as well as agent-based simulations.

We apply this methodological toolbox to two central dynamical processes of biological systems: (i) cell motility and motility control, and (ii) self-organized pattern formation inside cells and organisms.

Thereby, we seek to understand physical mechanisms that ensure robust biological function in the presence of non-equilibrium fluctuations, structural variations, and external perturbations.

## 1.3 Nonlinear dynamics and control

Active biological systems such as cells, tissues, and organisms continuously convert chemical energy into work and heat to facilitate *e.g.* directed motility and information processing[24]. Additionally, these systems are able to form ordered spatial patterns at the cell, tissue, and organism level by means of self-organization[20]. In both cases, nonlinear feedback loops control biological dynamics. This nonlinear control ensures robust function in the presence of fluctuations and perturbations. In addition to external perturbations, internal fluctuations arising from non-equilibrium molecular processes and small-number fluctuations in biochemical reactions can be substantial and impact the dynamics on mesoscopic scales.

**Nonlinear dynamics.** We can characterize biological systems in terms of mesoscopic variables. These variables may refer to classical biological variables such as the expression of specific genes or protein concentrations, as well as physical variables such as forces and fluxes, or spatial order parameters. A combination of positive and negative feedbacks between these mesoscopic variables gives rise to a rich nonlinear dynamics, whose features include excitability, bistability, and spontaneous oscillations[25–28]. These features enable responses to external stimuli and cellular decision making[27,29,30].

Excitability has been well characterized in the context of neuronal dynamics[25]. In publication 3.1, we will encounter an example of excitability in a pattern formation system. Bistability allows cells to dynamically switch between two cellular programs, *e.g.* modes of metabolic activity, in an adaptation to environmental conditions[31]. We will encounter an example of dynamic switching between two different steering modes in the context of chemotaxis of sperm cells in publication 2.4.

Oscillations are paramount in biology: they are observed *e.g.* in cellular signaling systems. In these systems, closed feedback loops with a temporal delay represent a generic design paradigm for spontaneous oscillations[32]. A well-studied example of a biological oscillator is the circadian clock, which sets day-night rhythms of biological activity[33]. In the circadian clock, signaling proteins regulate their own concentrations in a closed feedback loop with delays, resulting in oscillations with an intrinsic oscillation period of about $24\,\mathrm{h}$. These spontaneous oscillations become entrained to the daily rhythm of light exposure, providing an example of synchronization. Generally, signaling systems that harbor an internal oscillator can serve as a bandpass filter that actively amplifies oscillations of a sensory input signal at a certain frequency. An example is provided by hair cells of the inner ear that detect sound waves[34,35]. Some swimming cells process oscillatory light or chemical stimuli while navigating along chiral paths[7,8,36,37], which we address in publication 2.4.

Spontaneous oscillations occur also in chemo-mechanical oscillators. An important example are motile cilia and flagella, which represent slender cell appendages of eukaryotic (non-bacterial) cells[1–3]. Cilia and flagella exhibit self-organized regular bending waves with typical frequencies in the range of $10-100\,\mathrm{Hz}$. This flagellar beat pumps fluids and propels cellular swimmers in a liquid. Flagellar bending waves result from the collective dynamics of molecular motors inside the flagellum. A closed feedback loop, where elastic deformations of the flagellum control spatial profiles of motor activity inside the flagellum, gives rise to a dynamic instability and self-organized oscillations[38,39]. These chemo-mechanical oscillations do not depend on inertia, as motion is highly overdamped at the relevant length and time-scales. Instead, a combination of positive feedback and negative feedback with delay is sufficient to drive oscillations[32]. Positive feedbacks are a result of active processes and have been termed negative friction in the context of collective motor dynamics.

In addition to temporal dynamics, closed feedback loops also account for spatial patterns. Bidirectional, local interactions between two spatial fields $A(x)$ and $B(x)$ can give rise to the self-organized formation of spatial patterns[40,41], as discussed in section 1.3.2. These spatial fields can correspond to local concentrations of signaling molecules, or also local mechanical stress[40,42,43]. In publication 3.2, we will present a generic mechanism for self-organized pattern formation, whose patterns adapt to system size by nonlinear feedback control[11].

The nonlinear dynamics of biological system facilitates adaptation to external perturbations, and robust function in the presence of strong fluctuations. These are discussed in the following.

**Adaptation.** Cells and tissues can adapt to external conditions that change in time. A prototypical example is provided by sensory adaptation, where the sensitivity towards an extracellular stimulus is dynamically adjusted in response to slow changes of the stimulus base-level[15,17]. Sensory adaptation allows the detection of relative changes of a stimulus on a time-scale faster than a time-scale $\tau$ of adaptation. For sake of illustration, we consider a minimal model of sensory adaptation that has been abstracted by Barkai and Leibler from the more complex signaling dynamics of the chemotactic response of the bacterium *Escherichia coli* as[17]

$$\tau \dot{p} = 1 - ps. \qquad (1)$$

Here, a single nonlinearity, the product of the external stimulus $s(t)$ and a dynamic sensitivity $p(t)$, ensures that the system's output $a = ps$ is independent of the stimulus level for constant stimulus, $s(t) = s_0$, yet faithfully tracks relative changes of the stimulus on time-scales faster than $\tau$. We will employ an extension of this minimal model of sensory adaptation in a theoretical description of sperm chemotaxis in publication 2.4.

In addition to the dynamic adaptation of sensitivity levels, even functional spatial structures can adapt to external perturbations. For example, in cells that mechanically interact with an elastic substrate, the spatial organization of the cytoskeleton and its rheological properties change as a function of substrate stiffness[44,45] (see also theoretical work by the author on this topic[46–48]). On the tissue and organ level, examples of structural adaptation include the growth of muscle in response to exercise, or the thickening of bones in response to mechanical load[49,50]. Complex tissues such as the liver adapt to changes in metabolic load. These examples highlight the dynamic adaptation of form to function in biological systems. This dynamic adaptation requires a reverse feedback of functional characteristics on the structures that generated this function in the first place. Adaptation represents a specific case of information processing in biological systems. We now turn to another instance of cellular information processing, motility control, which offers the unique opportunity to directly observe the output of cellular signaling in the form of cellular motility responses.

**Motility control.** The control of cell motility requires closed feedback loops that link motility and sensory input. During chemotactic navigation of cells reviewed in section 1.4.3, external chemical stimuli are transduced by the cell to control the dynamics of the cytoskeleton of the cell, and thus cellular motility. Conversely, the active motion of a cell in a spatial field of a stimulus determines the temporal stimuli perceived by the cell. This general principle, by which a motile agent structures the sensory input it receives by its own motion has gained recent attention in the field of control theory as the principle of *information self-structuring*[51]. Steering responses of a cell represent a direct read-out of the output of the signaling cascades that control motility. Thus, cellular motility control represents a convenient model system to study the nonlinear dynamics of cellular information processing.

**Robustness.** An important aspect of dynamic feedback control in biological systems is the robustness of biological function to external perturbations and internal fluctuations[52,53]. At the mesoscopic scale of the cell, thermal noise, non-equilibrium fluctuations, and molecular shot noise can be substantial and interfere with biological function. For example, small-number fluctuations of signaling molecules introduce a substantial element of stochasticity into biological information processing. In section 1.4.3, we review three different chemotaxis strategies employed by single cells, each of which allows to detect extracellular concentration gradients of signaling molecules in a different way. We argue that these different mechanisms represent an adaptation to different levels of noise, both in motility and sensing[8]. Occasionally, fluctuations can play also a beneficial role: some cells harness noise to facilitate a spectrum of heterogeneous responses despite their otherwise identical setup, the most prominent example being the adaptive immune system[54]. Theoretical descriptions of biological function as pursued here allow to assess the reliability of control mechanisms with respect to external perturbations and intrinsic fluctuations. In addition to robustness with respect to intrinsic and extrinsic fluctuations, biological control designs often exhibit structural robustness.

Structural robustness defines the property of a system to function reliably, even if parameters of the system, or even its design, are varied. Such variability can be the result of genotypic heterogeneity, or of external perturbations and internal fluctuations that occurred during the development of the system. Control mechanisms that require fine-tuning of parameters would lack structural robustness. Theoretical descriptions of biological function allow to delineate the parameter region of reliable function. We will discuss examples of parameter robustness in chapters 2 and 3 in the context of motility control and pattern formation, respectively. Structural robustness relates also to the very design of the control mechanism itself. One common design paradigm for structural robustness is redundancy, where important functional elements operate in duplicate. Redundancy applies for example at the level of proteins, where several proteins often perform similar functions, and can partly substitute for one another, if one protein were absent. Similar, complex signaling networks often have redundant network topologies that can compensate for the failure of individual signaling links. Another design paradigm ensuring structural robustness are control mechanisms that depend only on qualitative features of functional relations between state variables (*e.g.* monotonic dependence of one variable on another) as compared to strict quantitative relations (*e.g.* linear dependence). In publication 3.2, we will explicitly discuss a generic pattern forming system that scales self-organized patterns proportional to system size, whenever a number of qualitative conditions are met[11].

In the following, we review selected aspects of nonlinear dynamics and feedback control with a focus on cell motility and self-organized pattern formation.

### 1.3.1 Mechanisms of cell motility

Cells employ a great variety of energy-dependent mechanisms for locomotion, including swimming, crawling, and twitching as discussed below[55]. A common feature of these different mechanisms is the non-equilibrium dynamics of the cytoskeleton of the cell[12]. Active shape changes allow motile bacteria and flagellated eukaryotic cells such as sperm to propel themselves in a liquid[56–58]. In these examples, molecular motors interact with cytoskeletal filaments to drive motility. Other cells such as macrophage immune cells crawl on a substrate by harnessing active polymerization forces of cytoskeletal filaments, which push their cell front forward[59,60]. This crawling motility requires partial adhesion to a substrate in order to constrain backward motion due to reaction forces.

Directed motion requires a structural polarity of the cell. Cell polarity can be static, as in the case of sperm cells with a defined head-tail morphology. Static cell polarity implies that cells have to actively rotate during steering responses. Other cells, such as macrophages with crawling motility, display a dynamic polarity, associated with a continuous remodeling of their cytoskeleton[61,62]. Multiple sensory cues including chemical and mechanical stimuli control the direction of cell motility.

**The molecular machinery of cell motility.** We first review key components of the cytoskeleton, whose non-equilibrium dynamics drives the different locomotion strategies of single cells: these components comprise structural biopolymeric filaments and force-generating molecular motors, see Figure 2.

*Three classes of cytoskeletal filaments in eukaryotic cells.* Inside cells, monomers of cytoskeletal proteins polymerize into filaments that constitute the cytoskeleton of the cell[3,63]. The cytoskeleton defines the mechanical properties and morphology of cells, especially in cells that lack a cell wall, such as animal cells. In eukaryotic cells, three classes of cytoskeletal filaments are found: actin filaments, microtubules, and intermediate filaments.

- *Actin filaments:* Actin is the most abundant intracellular protein in the eukaryotic (non-bacterial) cell, constituting 1-5% of its total protein content. Actin monomers (G-actin) polymerize into semiflexible actin filaments (F-actin), which have a persistence length of about $10\,\mu\mathrm{m}$ [63]. Actin filaments are structurally polar, with a designated plus-end (also named: barbed end) and minus-end (also: pointed end). In a typical eukaryotic cell, actin filaments form a crosslinked meshwork with gel-like properties that fills intracellular space. Additionally, actin filaments form a dense cortical network beneath the cell membrane of animal cells, the actin cortex. Turn-over of the actin cytoskeleton is fast, with a time-scale of $1-10\,\mathrm{s}$ measured for the actin cortex[64].

Polymerization dynamics of actin filaments is coupled to the hydrolysis of Adenosine triphosphate (ATP)[3,63]. This renders actin polymerization a non-equilibrium phenomenon that breaks detailed balance. Generally, polymerization kinetics is faster at the structural plus-end of an actin filament compared to its minus-end. Polymerizing actin filaments can exert active polymerization forces[65], which underlie the mechanism of crawling motility of cells[59].

The structure of the actin cytoskeleton is tightly regulated by the cell[3]. Specifically, the length of actin filaments is fine-tuned by capping proteins that cap filament ends to regulate actin polymerization dynamics. Severing proteins can bind at any position along an actin filament, inducing filament breakage at the binding position. Actin binding proteins can crosslink and bundle actin filaments. In addition to these 'passive' actin binding proteins, actin filaments interact with molecular motors of the myosin family that generate active forces[3]. The structural polarity of actin filaments with a designated plus- and minus-end sets a direction of motor motion. Conventional myosin motors walk towards the actin plus-end. Further, actin filaments, myosin motor proteins, and actin binding proteins can assemble into spatially ordered structures inside cells. For example, non-motile animal cells adhered to a substrate can form stress fibers of bundled actin filaments, thus representing a case of nematic order. In striated and cardiac muscle cells, actin filaments and myosin filaments are arranged in myofibrils of almost crystalline regularity, thus representing an example of smectic order of the cytoskeleton[22].

- *Microtubules:* The second major class of cytoskeletal filaments are microtubules, which are polymerized out of stable dimers of the protein tubulin. Microtubules are comparatively stiff hollow tubes of diameter $24\,\mathrm{nm}$ with a persistence length of about $1\,\mathrm{mm}$ [63]. Microtubules serve as tracks for kinesin and dynein motor proteins and play a major role in directed intracellular transport.

Microtubules can assemble into cell-scale ordered structures. One prominent example is the mitotic spindle, a bipolar cytoskeletal scaffold that serves for partitioning the two copies of the chromosomes to the two prospective daughter cells before cell division[3]. A second microtubule-based structure is the flagellar axoneme, which forms the cytoskeletal core of cilia and flagella. The axoneme comprises a cylindrical arrangement of 9 doublet microtubules, which are connected by dynein molecular motors (and additional proteins ensuring structural integrity)[66]. The collective dynamics of these motors drives regular bending waves of motile flagella[2], see Figure 4.

- *Intermediate filaments:* As a third class, intermediate filaments represent a heterogeneous family of filaments that serve as structural elements, *e.g.* in neurons and muscle cells. Special intermediate filaments form the hairs and nails of animals[3].

In bacteria, cytoskeletal filaments homologous to those of eukaryotic cells are found, which play important roles for cell motility and cell division[67].

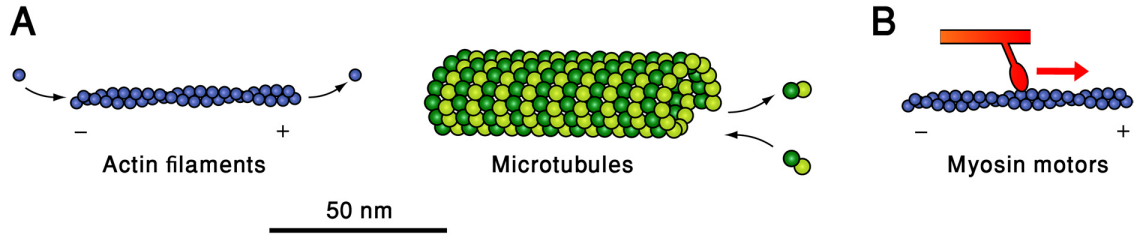

**Figure 2:** *Elements of the cytoskeleton.* **A.** Eukaryotic (non-bacterial) cells contain actin filaments and microtubules as key elements of their cytoskeleton, which defines mechanical properties of the cell. These biopolymers are highly dynamic and continuously undergo non-equilibrium polymerization dynamics. Actin filaments and microtubules are structurally polar, with distinct polymerization dynamics at their structural plus- and minus-end, respectively. **B.** Actin filaments and microtubules serve as tracks with defined directionality for molecular motor proteins, such as myosin motors. Myosin motors undergo chemo-mechanical cycles, which couple the energy-favorable hydrolysis of ATP molecules and a conformational change, which can generate piconewton forces and perform mechanical work.

**Non-equilibrium polymerization dynamics.** Polymerization of cytoskeletal filaments is a non-equilibrium process that is coupled to the hydrolysis of ATP in the case of actin filaments and GTP in the case of microtubules[3,63]. We briefly review non-equilibrium polymerization dynamics for the case of actin filaments[63]. Each actin monomer tightly binds either an ATP or ADP molecule. We thus refer to T-state and D-state monomers, respectively. Free monomers in the cytosol are mainly in T-state, while the monomers within an actin filament rapidly switch to D-state by hydrolysis of their bound ATP. An actin filament will elongate by polymerization at its tip, whenever the concentration of free monomers exceeds the critical concentration of the polymerization reaction. The critical concentrations for T-state and D-state monomers are different due to different values of $\Delta G$ for the respective polymerization reactions. For intermediate concentrations of free actin monomers, there can be net polymerization of T-state monomers at the structural plus-end of an actin filament, and net depolymerization of D-state monomers at the structural minus-end. As a result, a dynamic steady state can form that is characterized by net elongation at the plus-end and net shrinkage at the minus-end. During this actin treadmilling, actin monomers 'flow' through the filament. This mechanism requires that the rate at which new T-state monomers are added at the plus-end is faster than the rate of hydrolysis, such that the plus-end-tip of the filament will remain in T-form. The treadmilling of individual actin filaments captures essential aspects of crawling cell motility, which is driven by the non-equilibrium polymerization dynamics of a structurally polarized actin cytoskeleton[68]. In publication 3.1, we further discuss a possible role of actin treadmilling for the formation of periodic cytoskeletal patterns[10].

Actin filaments and microtubules serve as tracks for molecular motors, which we review in the next paragraph.

**Molecular motors convert chemical energy into work and heat.** Directed transport processes inside cells, cell locomotion, and contraction of muscle all rely on the activity of motor proteins at the molecular scale. The common working principle of a molecular motor is a tight coupling between an energy-favorable chemical reaction and a conformational change of the motor protein itself. This conformational change can perform mechanical work (with a typical order of magnitude of $1-10\,\mathrm{pN}\,\mu\mathrm{m}$ per chemo-mechanical cycle).

Different classes of molecular motors exist in bacterial and eukaryotic cells. In the cell membrane of bacteria such as *Escherichia coli*, a rotary motor is driven by a proton-gradient[56]. This rotary motor rotates helical filaments for cell propulsion[69]. In eukaryotic cells, molecular motors move along actin filaments and microtubules to transport cargo and generate active mechanical forces[63]. Important motor families include myosin motors, which move along actin filaments, and kinesins and dyneins, which move along microtubules. In their function as motor tracks, actin filaments and microtubules provide a periodic lattice of motor binding sites with a lattice constant of a few nanometers, which is set by the size of their respective monomers. The structural polarity of actin filaments and microtubules with a designated plus-end and minus-end defines a direction of motor motion. Most members of the kinesin motor family walk towards the structural plus-end of microtubules, whereas most dyneins walk towards the minus-end. Conventional myosin motors move towards the plus-end of microtubules. Molecular motors undergo periodic chemo-mechanical cycles, during which the motors bind and unbind from their track to take a single step, while one ATP molecule is hydrolyzed. We review this chemo-mechanical cycle for the example of skeletal myosin[63]: In the most common reaction path, a free myosin motor domain binds an ATP molecule, which is subsequently hydrolyzed into Adenosine diphosphate (ADP) and a phosphate group. The release of the reaction products constitutes the rate limit step of the ATPase activity of free myosin. Binding of myosin to an actin filament accelerates this release at least 200-fold. The release of ADP and phosphate is accompanied by a conformational change of the myosin motor domain, which causes a motion of the myosin backbone relative to the actin filament with a working distance of about $5\,\mathrm{nm}$. The myosin is then ready to bind a new ATP-molecule. This triggers the unbinding of myosin from the actin filament to restart the cycle.

Single molecular motors such as myosin and kinesin exert typical forces in the piconewton range. For example, conventional kinesin motors can exert forces up to $6\,\mathrm{pN}$, while taking $8\,\mathrm{nm}$ steps along their microtubule track. This corresponds to a mechanical work of $10\,k_BT$ per step. This represents a considerable fraction of the difference in Gibbs free energy of $\Delta G = 20-25\,k_BT$ associated with the hydrolysis of a single ATP molecule during each step[63]. Kinesin is a processive motor that takes a sequence of steps before it detaches from its track. Such processive molecular motors exhibit effective force-velocity relationships[70]: an applied external force reduces their velocity, until their motion comes to a halt at a critical stall force. For conventional kinesin, the stall force is about $6\,\mathrm{pN}$.

We note that for a single molecular motor, the principle of microscopic reversibility holds: for each reaction step of the chemo-mechanical cycle, both the forward and the backward reaction are possible. Thus, there is a finite probability that a molecular motor takes a step backwards. At physiological conditions, the high chemical potential of ATP breaks detailed balance of the cycle and favors forward motion. Backward stepping of molecular motors has been observed experimentally, especially under high load forces. As a side note, ATP-synthesis in mitochondria relies exactly on this microscopic reversibility: the $F_0$-$F_1$-ATPase protein complex couples a proton-driven and an ATP-driven rotary motor. In the presence of a strong proton gradient generated by glycolosis across the mitochondrial membrane, the proton-driven $F_0$ motor spins the ATP-driven $F_1$-motor backwards[71,72]. As a result, the $F_1$-motor serves as a dynamo that synthesizes its own fuel in the form of high-energy ATP molecules.

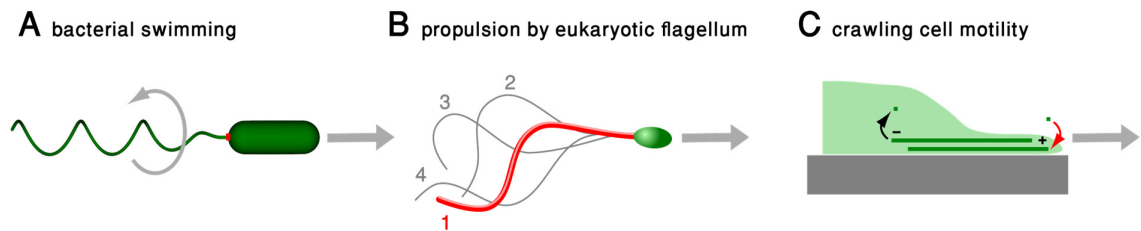

**Figure 3:** *Mechanisms of cell motility.* **A.** Bacteria such as *Escherichia coli* propel themselves in a liquid by rotating a passive helical filament, the prokaryotic flagellum. The rotation of this prokaryotic flagellum is driven by a rotary motor in the cell membrane, which draws its energy from a proton gradient across the cell membrane. Some bacterial strains are multi-flagellated with several prokaryotic flagella that can synchronize their rotations and form stable bundles. **B.** Eukaryotic (non-bacterial) cells such as sperm cells can swim in a liquid by virtue of regular bending waves of one or several eukaryotic flagella. Eukaryotic flagella are active filaments. Their bending waves emerge from the collective dynamics of a large number ($10^4 - 10^5$) of molecular dynein motors distributed along the length of the eukaryotic flagellum. **C.** Eukaryotic cells, such as macrophages of the immune system, harness polymerization forces of numerous actin filaments to crawl on a substrate. Propagation of a leading front termed lamellipodium is driven by polymerization forces of a structurally polarized actin cytoskeleton. Additional motility mechanisms are mentioned in the text.

**Bacteria swim by rotating passive helical filaments.** One of the best-studied examples of cell motility is the swimming of the bacterium *Escherichia coli*. This bacterium employs a rotary molecular motor in its cell wall to rotate a passive helical filament, termed the prokaryotic flagellum[56]. The prokaryotic flagellum is physically connected by a flexible hook to the rotor of the motor complex, which in turn can rotate freely inside a stator that is anchored to the cell wall. A proton gradient across the cell membrane drives a counter-rotation of rotor and stator[73]. This rotary motor has been a model system of biological physics, and its macro-molecular structure and mechanical function have been studied in great detail. The rotation of the rotary motor spins the helical filament and thereby propels the bacterium in a liquid[69], see Figure 3.

*Bacterial motility control.* Bacterial swimming represents a model system of motility control that has been studied extensively at the level of individual motors, of individual filaments, and at the level of the cell. Classic experiments revealed an operational load characteristic of the rotary motor with a rotation frequency that decreases with the applied load[74]. Such force-velocity relationships represent a general characteristic of molecular motors.

The prokaryotic flagellum itself is a passive filament. It is polymerized out of a single type of monomer, the protein flagellin. The prokaryotic flagellum forms a tubular polymer with 11 protofilaments. The helical shape of this filament is the result of a cooperative conformational change of flagellin monomers within a defined sub-set of its protofilaments. This heterogeneous conformational switch minimizes an intrinsic eigenstrain of the protein lattice in the flagellum[75–78]. Mechanical load can induce a cooperative conformational switching of all flagellins in one protofilament and thus a dynamic transition of the entire filament to a different polymorphic helical state. Most of the 11 theoretically possible polymorphic states have been observed in experiments.

In bacteria with multiple flagellar filaments, hydrodynamic interactions between rotating helical filaments results in the synchronization of filament rotation of the different filaments and the formation a stable bundle[79,80]. This flagellar bundling enhances propulsion efficiency. During bacterial swimming and navigation, flagellar bundling is tightly controlled by an intracellular signaling pathway[69]. Specifically, chemical signals can reverse the rotation direction of one or several rotary motors, which destabilizes the flagellar bundle and induces a transition of one or several flagella to a different polymorphic state of different handedness. The net result of this transient dynamics is a random reorientation event of the cell. A dynamic regulation of the frequency of these stochastic reorientation events facilitates chemotactic navigation in chemical gradients along a 'run-and-tumble' biased random walk[81]. This bacterial chemotaxis strategy is discussed in more detail in section 1.4.3.

While the bacterial flagellum is a passive filament, eukaryotic (non-bacterial) cells employ active filaments, termed cilia and flagella, which we discuss next.

**The eukaryotic flagellum is an actively bending filament.** Many eukaryotic (non-bacterial) cells are equipped with slender cell appendages termed cilia or flagella[1]. Cilia and flagella perform multiple sensory, signaling, and motility functions[82,83]. We will use the term eukaryotic flagellum for both cilia and flagella (where the main difference between cilia and flagella are their length and minor structural details). The eukaryotic flagellum is not to be confused with the prokaryotic flagellum of bacterial cells. While the prokaryotic flagellum is a passive protein polymer, the eukaryotic flagellum is an active filament.

The eukaryotic flagellum is a membrane-enclosed cell appendage of typical length $10-100\,\mu\mathrm{m}$ and diameter of about $500\,\mathrm{nm}$ that contains a highly regular cytoskeletal core, the axoneme[3], see Figure 4. The axoneme is composed out of 9 doublet microtubules in equidistant cylindrical arrangement, see figure 3. Additionally, a central pair of microtubules in the center of this cylinder may be present or not, corresponding to the sub-types of 9+2 and 9+0-axonemes. More than 250 accessory proteins ensure structural integrity and function of the axoneme[66]. The axoneme of motile eukaryotic flagella contains dynein motors[66,84], which render the eukaryotic flagellum a mechanically active filament[85].

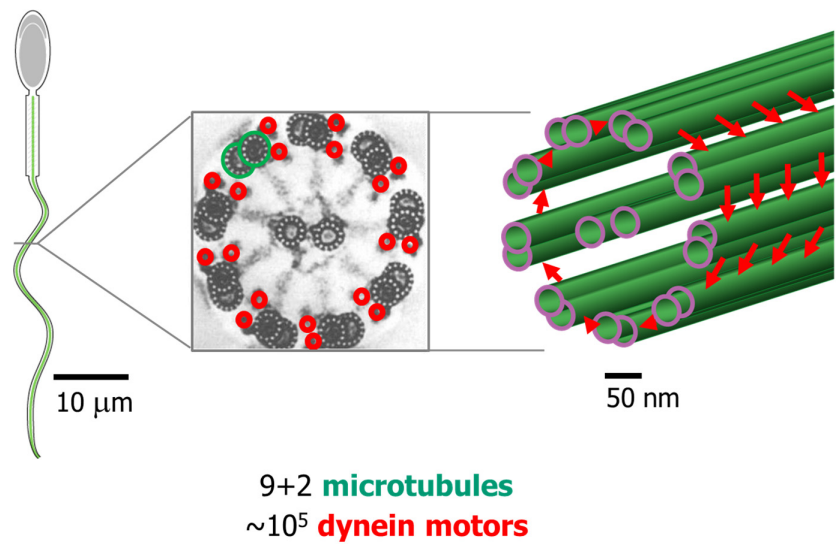

9+2 **microtubules**
~$10^5$ **dynein motors**

**Figure 4:** The eukaryotic flagellum contains a highly conserved cytoskeletal core, the axoneme. The axoneme comprises a cylindrical arrangement of 9 doublet microtubules, which are connected by dynein molecular motors. The collective dynamics of these dynein motors drives regular bending waves of cilia and flagella. *Left:* Schematic of flagellated sperm cell, *middle:* cross-section of the flagellar axoneme, *right:* schematic of axonemal architecture. Electron micrography from ref.[86] with permission.

It is remarkable that the highly regular structure of the axoneme is found in all 5 kingdoms of eukaryotic life, including amoeba, plants, and animals[87]. This evolutionary highly conserved structure appeared early after the chiasm between prokaryotes (bacteria and archaea) and eukaryotes. It has been speculated that the axoneme evolved from the cytoskeletal cell division machinery in eukaryotic cells, the mitotic spindle, by means of re-dedication to a new function[87].

**Collective motor dynamics drives flagellar bending waves.** Some cilia and flagella are motile. Inside their axonemes, neighboring doublet microtubules are connected by dynein molecular motors[66,84]. The axoneme has a chiral architecture: dyneins are tightly bound to one doublet and exert forces on the neighboring doublet in clockwise sense (when viewed from the basal end of the axoneme).

We review the mechanism of active flagellar bending by motor-induced filament sliding[2,3]. The activity of dynein motors slides neighboring doublet microtubules relative to each other[88,89]. Free sliding is partially constrained, both at the basal end of the axoneme as well as by nexin protein links distributed along the flagellar length. These constraints convert the shearing forces generated by the dynein motors into bending moments that bend the axoneme. Bending in one direction requires that motors on one side of the axoneme are preferentially active at a given time. Spontaneous oscillations in motor activity drives regular bending waves of the flagellum. The bending rigidity of the flagellum is highly anisotropic for many flagella[90], favoring bending in a plane. This results in planar flagellar beat patterns in many cells, including important flagellated model swimmers such as marine invertebrate sperm or the green alga *Chlamydomonas*. A small chirality of flagellar bending waves results in helical swimming paths of defined handedness of these flagellated swimmers[91–93], which has implications for cellular navigation[7–9,94].

The control of dynein activity and the emergence of oscillatory motor activity represents an instance of self-organized collective dynamics in an ensemble of molecular motors[95]. One of the most-striking experiments demonstrating this self-organization is the re-activation of demembranated axonemes isolated from flagellated cells[96–98]. Upon provision of ATP, these isolated axonemes resumed regular bending waves, independent of any cellular control circuits.

Self-organized flagellar bending waves are the result of a closed feedback loop between the spatial activity profile of dynein motors inside the axoneme and geometric deformations of the axoneme, which gives rise to a dynamic instability[99,100]. Specifically, local motor activity deforms the axoneme, which again changes motor activity in a defined spatial range. As a result, travelling waves of motor activity emerge, which propagate along the flagellar length[39,100–102]. The shape of the resultant flagellar bending waves is sensitive to boundary conditions[100]. We have been general in refereeing to the geometric deformation of the axoneme on purpose. The precise nature of the control of motor activity by deformations of the axoneme is still open. Three major theories are discussed. In one of the earliest theories, Brokaw proposed that the local curvature of the axoneme constitutes the key regulator of motor activity[99]. Other authors objected that the local deformation resulting from typical curvatures are negligibly small on the length-scale of individual molecular motors[39]. Lindemann *et al.* proposed that bending of the axoneme causes splay, *i.e.* an increase of the inter-doublet spacing, which potentially could regulate motor activity[102]. Finally, Jülicher *et al.* considered a theoretical description in which the local sliding displacement of neighboring microtubules controls motor activity[39]. While the last model could quantitatively account for the waveform of the sperm flagellar beat, recent experiments with shorter flagella of the green alga *Chlamydomonas* challenges the sliding control model[103,104]. It is possible that control mechanisms of the flagellar beat are less conserved as previously anticipated. A major bottle-neck in uniquely identifying the mechanism of motor control of the beating axoneme is the simplicity of flagellar bending waves, which can be characterized by a small number of waveform parameters. Thus, different mechanisms relying on different microscopic assumptions can reproduce the observed waveforms equally well, provided the parameters of these models are chosen appropriately. Our research presented in this thesis provides additional characterizations of the flagellar beat in terms of (i) an active mechano-response of the flagellar beat in response to changes in hydrodynamic load[5] and (ii) active fluctuations of the flagellar beat due to motor noise[14]. We anticipate that such additional characterizations can contribute to the discrimination between the different proposed theories on the origin of the flagellar beat.

**Moving in fluids.** Flagellar bending waves propel cellular swimmers such as sperm cells[1,58], swimming alga[105], and pathogens (*e.g.* Trypanosomes[106], which cause sleeping sickness, and certain life cycle stages of Malaria parasites[107]). At the length and time-scales of cellular microswimmers, viscous forces dominate over inertial effects[108–110]. The relative magnitude of inertial forces compared to viscous forces for a swimmer with periodic shape dynamics is characterized by the dimensionless Reynolds number of oscillatory motion

$$\mathrm{Re} = \frac{\rho \omega_0 A d}{\eta}. \tag{2}$$

Here, $\rho$ and $\eta$ denote the density and dynamic viscosity of the fluid, respectively, while $\omega_0$ and $A$ denote frequency and amplitude of the periodic swimming stroke. Finally, $d$ denotes a characteristic length-scale of the swimmer. A low Reynolds number implies that viscous forces dominate over inertial forces at the relevant time and length-scales. For example, for a beating flagellum of diameter $d = 0.4 \, \mu\mathrm{m}$, beat amplitude $A = 5 \, \mu\mathrm{m}$, beat frequency $\omega_0 = 30 \, \mathrm{Hz}$, we estimate $\mathrm{Re} \sim 10^{-4}$. Note that it is the diameter of the flagellum, not its length, that sets the magnitude of maximal fluid stresses[111].

In the limit of zero Reynolds number, the Navier-Stokes equation governing fluid flow simplifies to the linear Stokes equation,

$$0 = \nabla p - \eta \nabla^2 \mathbf{v}, \tag{3}$$

where $p$ and $\mathbf{v}$ denotes pressure and flow field of the fluid. The Stokes equation is linear. Thus, its solutions obey a superposition principle. General solutions of the Stokes equation can be found as superposition of its fundamental solution, the Stokeslet $v_i = G_{ij} F_j$, which describes the flow resulting from a point force $F_j \delta(\mathbf{r})$ acting on the fluid. Here, $G_{ij}(\mathbf{r}) = (8\pi\eta)^{-1}(1/r + r_i r_j/r^3)$ denotes the Oseen tensor. This superposition principle has been exploited to derive analytical results for the motion of minimal model swimmers, see *e.g.* references[112–116]. This superposition principles further underlies efficient algorithms to solve the Stokes equation in complex geometries numerically[117].

The second general feature of the Stokes equation is its invariance under time-reversal. This time-reversal symmetry has important functional consequences for swimming and hydrodynamic synchronization at low Reynolds number. Time-reversibility implies that the swimming path of a low Reynolds number swimmer depends only on the sequence of shapes it attains as a function of time, but is independent of the rate of shape change. In particular, a reciprocal swimming stroke with a backward stroke that traces a forward stroke exactly backwards in time will produce a periodic motion, but zero net displacement. This phenomenon is known under the colloquial name of the *Scallop theorem*[108]. The name was coined following an illustrative example of an idealized scallop presented in a popular lecture by Purcell: a two-leg-swimmer with a single joint that opens and closes periodically. A corollary of the scallop theorem is that any swimming stroke with amplitude $A$, where $A$ is small compared to a size $L$ of the swimmer, will result in a net swimming speed $\overline{v}$ that scales quadratically with $A$, *i.e.*

$$\overline{v} \sim A^2 \omega_0 / L. \tag{4}$$

Here, $\omega_0$ is the frequency of the swimming stroke. One may denote this relation the *quadratic law of low Reynolds number propulsion*[115]. Formally, the net swimming speed $\overline{v}$ is independent of fluid viscosity $\eta$. However, this argument assumes that the swimming stroke will not be altered by an increase in hydrodynamic load associated with an increase in $\eta$. In real systems, the active processes that drive the swimming stroke will generally display a force-velocity relationship, *i.e.* slow down under increased load. Such force-velocity relationships have been measured for beating flagella both in response to an increase in fluid viscosity[96,118,119] as well as in response to a dynamically varying load[5].

The mathematical beauty of self-propulsion at low Reynolds numbers has attracted a continuous stream of theoretical studies, following an early exposition by Taylor in 1951[4]. We briefly mention a geometric interpretation by Shapere and Wilczek, who identified self-propulsion at low Reynolds numbers as a connection of an $SE(3)$-fiber bundle on a space $S$ of admissible shapes of a swimmer[120]. Here, the special Euclidean group $SE(3)$ denotes the Lie group of rigid body transformation in 3-dimensional space. Any trajectory $s(t)$ in shape space $S$ lifts to a trajectory $[s(t), \mathbf{G}(t)]$ in this fiber bundle $S \times SE(3)$. This trajectory characterizes the translational and rotational motion of a shape-changing low Reynolds number swimmer that is free from external forces and torques in terms of a time-dependent rigid body transformation $\mathbf{G}(t) \in SE(3)$ of a material frame of the swimmer. For periodic shape changes of small amplitude $A$, the net motion $\mathbf{G}(T)^{-1}\mathbf{G}(0)$ after one swimming stroke of period $T$ is proportional to $A^2$, see equation (4). The proof relies on the argument that any swimming stroke of infinitesimal amplitude can be written as the superposition of several reciprocal shape modes. As a consequence of the scallop theorem, none of these reciprocal shape modes alone can result in any net motion of the swimmer. However, nonlinear cross-terms between different reciprocal modes result in net displacement, which thus scales with $A^2$. As an example, we note that flagellar bending waves can be approximately described as traveling bending waves. A traveling bending wave can be written as the superposition of two standing waves phase-shifted by $\pi/2$ by elementary trigonometry. Each standing wave alone would provide zero net propulsion, while their superposition allows for flagellar self-propulsion[121].

We conclude that the time-reversal symmetry of the Stokes equation prompts non-reciprocal swimming strokes that explicitly break time-reversal symmetry to allow for net propulsion. Similarly, we will find that hydrodynamic synchronization at low Reynolds numbers is only possible if specific symmetries are broken[110,115,122], see also section 1.4.2.

**Flagellated microswimmers represent a model system for motility control.** Cilia and flagella are a best-seller of nature. Virtually all animal cells display one or more of these slender cell appendages, which serve for motility and sensing[1,123]: non-motile primary cilia facilitate our senses of smell, vision, and, in some species, hearing, and gauge blood flow to regulate blood pressure. Motile cilia and flagella propel sperm cells, green algae, and disease-causing protists, such as Trypanosomes (responsible for sleeping sickness) or plasmodia (which cause malaria) in a fluid[124]. Sensory input controls flagellar beating in these microswimmers and enables them to actively steer their path in response to environmental cues[7–9]. On epithelial surfaces, carpets of short flagella termed cilia synchronize their beat to pump fluids, such as mucus in mammalian airways[125] and cerebrospinal fluid in the brain[126]. Chiral flagellar beating plays a crucial role in the establishment of the left-right body axis during embryonic development[127,128].

**Flagellar swimming, steering, and synchronization.** The flagellar beat − itself a manifestation of microscopic dynamics of dynein motors inside the flagellar axoneme − facilitates swimming and steering in flagellated microswimmers. The asymmetric shape of the flagellar beat determines the chiral swimming paths of these cellular swimmers. A dynamic regulation of beat shape underlies steering in response to chemical signals, light, and possibly temperature. Finally, speed and shape of the flagellar beat are susceptible to external mechanical forces. This flagellar load response is a prerequisite for the remarkable phenomenon of beat synchronization by mechanical coupling in collections of beating flagella.

*Flagellar bending waves are chiral.* In sperm cells and green alga, flagellar beat patterns resemble planar bending waves that propagate along the flagellum, from its proximal to its distal tip[1]. These flagellar bending waves commonly display a pronounced in-plane asymmetry, characterized by a static mean curvature of the flagellar shape. This mean flagellar curvature $K_0$ is a result of active processes inside the flagellum. Experiments with reactivated flagellar axonemes revealed that $K_0$ depends on ATP concentration[98]. This finding is consistent with the notion that flagellar asymmetry is generated by static motor activity inside the flagellum. Changes in the viscosity of the surrounding fluid reduced the mean flagellar curvature $K_0$ [119]. During chemotactic steering responses of sperm cells, an intraflagellar signaling cascade dynamically regulates flagellar asymmetry $K_0$ [129,130]. The microscopic origin of flagellar beat asymmetry remains insufficiently understood.

In addition to the static and dynamic component of flagellar curvature, the flagellum is also twisted. This results in non-planar beat patterns that break chiral symmetry[93,131,132]. Flagellar twist is small in the green alga *Chlamydomonas*, as well as the beat of the flagellar beat in sperm of many model species. In mouse and humans, a pronounced flagellar twist gives rise to conical flagellar waves[133]. Flagellar twist is also pronounced for cilia on epithelial surfaces, whose beat pumps fluids. These cilia move backward during their recovery stroke in close proximity to the surface in a highly twisted configuration.

*The chiral flagellar beat controls swimming and steering.* Chiral flagellar beat patterns result in chiral swimming paths of flagellated swimmers such as circles, twisted ribbons, and helices[93,119,134]. These chiral swimming paths form the basis of dedicated sampling strategies of cellular navigation[7–9], which are reviewed in section 1.4.3. Far from boundary surfaces, sperm from marine invertebrates swim along helical paths[91–93]. From hydrodynamic simulations, we could estimate the flagellar twist required to account for the observed helicity of swimming paths[93]. This flagellar twist is surprisingly small and corresponds to an out-of-plane component of the planar flagellar bending waves of less than a micrometer. In the vicinity of boundary surfaces, sperm cells localize close to the interface[135], where they swim in circles with defined sense of rotation relative to the surface normal[119]. It has been proposed that the chirality of the flagellar beat contributes to this surface accumulation of sperm in addition to pure hydrodynamic effects[136–138].

Swimming of sperm cells along circular and helical paths represents a stereotypic form of exploratory motion, which forms the basis of a dedicated navigation strategy. Sperm from marine species perform helical chemotaxis[8,93,139–142] to steer their swimming path up chemical gradients of signaling molecules to find the egg[143,144]. In the unicellular green alga *Chlamydomonas*, a small out-of-plane component of the beat of the its two flagella causes a self-rotation of the cell around its swimming axis at a frequency of about $2\,\text{Hz}$ (which is much slower than the beat frequency of $50\,\text{Hz}$)[105]. This slow rotation allows this green alga to detect directed light stimuli as cell rotation periodically exposes a light-sensitive eye spot to incident light. A simple steering feedback allows these cells to steer their swimming path relative to the direction of incident light[9,94,145].

Finally, during embryonic development, the chiral beat of flagella generates leftward flow of fluid, which determines the left-right-body axis in the developing embryo[146]. The broken chiral symmetry of the flagellar beat results neither from a spontaneous symmetry breaking, nor does it depend on physical laws that explicitly break chiral symmetry. Instead, the chiral flagellar beat is a result of the chiral architecture of the flagellar axoneme[131], which in turn is rooted in the chirality of the proteins it is built of. Chiral flagellar beat patterns represent an example of homo-chirality that propagates from the molecular scale all the way up to the scale of cells, tissues and organisms.

**Mechanical control of flagellar motility**. The beating flagellum exerts active forces on the surrounding liquid. Conversely, external mechanical forces affect the swimming path of flagellar microswimmers. Additionally, external forces change the speed and shape of the flagellar beat itself. This mechanical control of the flagellar beat is important for flagellar mechano-taxis and flagellar synchronization in collections of flagella, as reviewed below.

*Interactions with flows and structures.* Fluid-structure interactions control flagellated motility in a number of functional contexts. It had been noticed already by Rothschild that sperm cells localize near a glass-water interface, where sperm cells swim in a plane[135]. This phenomenon can be explained as a pure hydrodynamic effect. The chirality of the flagellar beat of sperm cells induces a thrust component normal to the interface that brings cells closer to the interface until a critical distance is reached, where short range repulsion sets in[136,138]. At least two additional hydrodynamic effects stabilize swimming at the interface. First, the leading order singularity of the time-averaged flow field generated by a swimming sperm is that of a pusher, with an inward flow component along the normal of the beat plane. This induces a hydrodynamic attraction the wall[109]. Second, swimming at a small distance $d$ to the wall, which is less than the flagellar length $L$, effectively suppresses rotational diffusion of the cell, thereby prolonging residence times close to the interface[137].

A second instance of flagellar motility control by mechanical forces is the active upstream migration of sperm cells, termed rheotaxis[147,148]. In the presence of walls, local shear flows rotate swimming sperm like a weather van such that the vector of their net swimming direction points upstream[149]. It has been proposed that this rheotaxis serves as a sperm guidance mechanism in the mammalian oviduct, where post-coitus oviductal flows are sufficiently strong to align sperm[148].

As a final example of microswimmer-structure interactions, flagellated trypanosomes were found to swim more efficiently in the presence of obstacles, whose size matches the radius of curvature of flagellar bending waves[150]. This has been interpreted as an adaption to the crowded environment of the blood stream, where red blood cells represent semi-rigid obstacles.

*Active mechano-responses.* Self-organized flagellar bending waves exhibit active mechano-responses with a flagellar wave form that depends on applied external forces. Early experiments by Brokaw have shown that an increase in the viscosity of the swimming medium reduces both the frequency and the amplitude of the flagellar beat[96,118]. Similar experiments using local micropipette-generated flows or swimming sperm in a visco-elastic fluid gave qualitatively similar results[151,152].

The flagellar beat is an emergent phenomenon of collective dynamics in an ensemble of dynein molecular motors working against both intra-flagellar forces and hydrodynamic friction forces, which result from moving the surrounding fluid. It is thus to be expected that changes in these external forces change the shape of the flagellar beat. The waveform compliance of the flagellar beat provides a rough estimate of the relative importance of intraflagellar friction forces and hydrodynamic friction forces. In publication 2.2, we theoretically derive a force-velocity relationship of the flagellar beat[5]. The corresponding effective theory coarse-grains active motor dynamics in the flagellar axoneme in terms of a phase-dependent active driving force. Our theoretical predictions are compared to a dynamic measurement of this force-velocity relationship in the green alga *Chlamydomonas*. We provide an analysis of experimental data that shows how the phase speed of the flagellar beat changes in response to rotations of the cell, which imparts known hydrodynamic friction forces on the flagellum, allowing us to infer how the beating flagellum responds to external fluid forces.

Recent experiments suggest that the ATP consumption of flagellar beating is rather insensitive to mechanical load[153]. This experimental finding resonates with a theoretical description that employs a fixed phase-dependent flagellar driving force to represents the active dynamics inside the axoneme that makes the flagellum beat. Accordingly, any increase in load is compensated by a reduction in speed, not an increase in fuel consumption.

For single processive motors such as kinesin force-velocity curves have been measured in single-molecule experiments[70]. Analogous measurements for non-processive axonemal dynein are not known, let alone their collective dynamics. Analysis of the flagellar beat under different load conditions provides a means to measure this force-velocity relationship.

*Flagellar synchronization.* The force-velocity relationship of the flagellar beat is an essential prerequisite for the striking phenomenon of beat synchronization by mechanical coupling. It had been observed already by Gray almost 100 years ago that pairs of sperm cells swimming in close proximity can synchronize their flagellar beat[1,154]. Similarly, sperm held in vibrating micropipettes or exposed to oscillatory flows entrain to the frequency of external driving[155,156]. Flagellar synchronization plays an important role for the collective dynamics in ciliar carpets, for example on the surface of unicellular *Paramecium* or the epithelial surfaces of mammalian airways[125]. Emergent metachronal waves enable fast swimming and efficient fluid pumping[127,157,158]. It has been proposed already by Taylor in 1951 that a mechanical coupling between several flagella can synchronize their beat[4]. Obviously, such a mechanical coupling requires a dependence of the speed of the flagellar beat on mechanical forces. In publication 2.2, we could identify a mechanism of flagellar synchronization in free-swimming *Chlamydomonas* cells. The unicellular green alga *Chlamydomonas* swims like a breast-stroke swimmer with two flagella that can beat in synchrony[5,159–161]. Synchronized beating is important for *Chlamydomonas* to swim fast and straight[5,162]. Flagellar synchronization relies on a force-velocity relation of the flagellar beat, which we predict theoretically and characterize experimentally by analyzing experimental data[5].

Mutual interactions between flagellated swimmers can even give rise to dynamic pattern formation in collections of microswimmers. In dense suspensions of sperm cells, swimming in circles close to a boundary surface, sperm organized into vortices of about 5 cells each[163]. Inside each vortex, sperm flagella phase-locked their individual flagellar waves into a fixed phase-relationship. Vortices organized into regular hexagonal patterns, presumably due to an effective repulsion between neighboring vortices.

**Chemical control of flagellar motility.** During flagellar swimming, the shape of the flagellar beat is under tight control of intracellular signaling. This chemical beat control facilitates in particular flagellar steering responses during chemotaxis[129,130,164], phototaxis[94,165], and mechanotaxis[166]. For example, sperm cells from marine invertebrates dynamically regulate the asymmetry of their flagellar beat to steer their path up a concentration gradient of chemoattractant[129,130]. The bi-flagellated green alga *Chlamydomonas* can switch from normal forward swimming to backward swimming upon exposure to strong, potentially harmful light stimuli[7,94,165]. During these photoshock responses, the asymmetry of the flagellar beat is greatly diminished in both flagella. In weak light conditions, a differential regulation of beat amplitude causes a yawing motion of the cell to facilitate phototaxis towards the direction of incoming light[165]. Mammalian sperm cells switch from travelling flagellar bending waves to a state of vigorous flagellar motility termed 'hyperactivation' during a ripening process inside the oviduct[143,167].

*The intraflagellar calcium is a key regulator of the flagellar beat.* Classic experiments by Brokaw in reactivated, demembranated axonemes have shown that an increase of calcium concentration increased the asymmetry of the flagellar beat in a gradual manner[168]. Studying the relationship of intraflagellar calcium and beat asymmetry in intact flagella is challenging, as methods for simultaneous manipulation and monitoring of calcium concentration are required. Recently, sea urchin sperm cells served as a versatile experimental model to address this question. Specifically, changes in intraflagellar calcium were evoked by activation of the chemotactic signaling cascade. Intraflagellar calcium concentration can be monitored using calcium-dependent fluorescent dyes. In these experiments, the instantaneous curvature of sperm swimming paths has been used as a proxy for a time-dependent flagellar asymmetry. These dynamic measurements revealed a dynamic relationship between fluorescence signal and path curvature that was well approximated by a smooth time-derivative[169]. This finding suggests that the transfer function from intraflagellar calcium to beat asymmetry is a band pass filter. Such a control design would serve as an additional layer of sensory adaptation in the chemotactic control of the flagellar beat during chemotaxis[8]. Measurement of fluctuations of the flagellar beat in the green alga *Chlamydomonas* revealed unusually long correlation times of seconds, which is much longer than the chemo-mechanical cycle times of molecular motors[151]. It has been proposed that these slow flagellar fluctuations are caused by fluctuations of intraflagellar calcium concentration.

It is not known by which molecular mechanism intraflagellar calcium regulates the shape of the flagellar beat. Axonemal dynein has several calcium binding domains and it is likely that calcium regulates motor activity[170]. Alternatively, it has been proposed that calcium control is indirect, mediated by calcium binding proteins such as calmodulin or calaxin[171,172]. Interestingly, *in vitro* motility assays with reactivated dynein suggests an indispensable role of calaxin for motor control[172]. In addition to calcium, the shape of the flagellar beat is also regulated by cAMP concentration and possibly pH[173].

**Other motility mechanisms of single cells.** We will only briefly mention alternative motility mechanisms employed by single cells. Despite their diversity, all these mechanisms rely on the non-equilibrium dynamics of the cellular cytoskeleton in one way or the other.

*Bacterial twitching motility.* Some bacteria use depolymerization forces for locomotion. For example, bacteria, such as *Neisseria gonorrhoeae*, extends passive protein filaments termed type IV pili, which adhere to neighboring cells or a substrate[174]. Active depolymerization at the cell-facing end of these pili generates forces of up to $100\,\mathrm{pN}$, rendering type IV pili the strongest molecular machines characterized so far[175,176]. Active pili retraction pulls the cells forward at speeds $v \lesssim 1\,\mu\mathrm{m\,s^{-1}}$ [174,177]. This twitching motility of bacterial cells plays an important role in biofilm formation.

*Crawling motility of eukaryotic cells.* Crawling motility of eukaryotic cells on a substrate depends to a large extent on polymerization forces. This locomotion strategy is employed *e.g.* by immune cells such as marcophages, and has been extensively studied in the model organism *Dictyostelium*, a slime mold[60,178]. These cells comprise a cytoskeleton of cross-linked actin filaments. Polymerization of actin monomers into actin filaments is a non-equilibrium process. Notably, net polymerization can occur even when the plus-end of the filament is pushing against an obstacle such as the cell membrane. In this case, each elongation of the filament by one monomer performs mechanical work[65]. The concerted action of an ensemble of polymerizing actin filaments results in a net forward propagation of the cell front, the lamellipodium. This crawling motility requires a self-polarization of the actin cytoskeleton that establishes a structural polarization of the cytoskeleton, to set the direction of motion. Signaling cues can bias the polarization direction, *e.g.* during chemotaxis of cells in external chemical concentration gradients[61].

After this short review of active cell motility and its dynamic control, we now turn to self-organized pattern formation as a second instance of spatio-temporal biological dynamics.

### 1.3.2 Self-organized pattern formation in cells and tissues

In systems far from equilibrium, such as living matter, local interactions between constituents can give rise to large-scale ordered patterns that represent dynamic steady states. Examples range from self-organized pattern formation inside cells, *e.g.* in the cytoskeleton, to the robust development of complex tissues and organisms with specific spatial order adapted to their function[179]. In Chapter 3, we will present two publications that address self-organized pattern formation at the cytoskeleton and at the organism level, respectively.

**Self-assembly of macro-molecular structures.** Inside cells, functional marco-molecular complexes self-assemble from individual molecules such as interacting proteins. Self-assembly can be passive as for the bacterial rotary motor, where more than 20 different proteins assemble stator and rotor of the motor complex[73,180]. Self-assembly of larger structures often involves active processes. For example, for the assembly of the prokaryotic flagellum, an active excretion system exports flagellin monomers, inserting them at the proximal end of the hollow flagellum, from where the monomers diffuse to the assembly site at the distal end[181]. The axoneme of the eukaryotic flagellum contains a bidirectional transport system known as intra-flagellar transport[182,183]. Therein, kinesin motors transport cargo, including the proteins that built the axoneme, towards the distal tip of the axoneme, whereas mobile dynein motors transport cargo back towards the proximal end, resulting in a stable steady state.

The interaction of cytoskeletal filaments and molecular motors gives rise to a variety of pattern formation phenomena, both *in vitro* and *in vivo*. This includes bundles of nematically aligned actin filaments, asters, stable swirling patterns[184–186], and even 'artificial cilia'[187]. A case of almost crystalline order of the cytoskeleton is found in myofibrils in striated muscle cells and cardiomyocytes[3]. Myofibrils are acto-myosin bundles characterized by a periodic arrangement of actin filaments of defined structural polarity and bipolar myosin filaments, which are organized in sarcomeric unit cells, see figure 5-C. In publication 3.1, we present a minimal mechanism for the self-assembly of periodic cytoskeletal patterns as observed in myofibrils[10]. This minimal mechanism relies on active force generation, such as active polymerization forces of treadmilling actin filaments.

**Pattern formation in ensembles of active particles.** Self-organized dynamic patterns naturally evolve in suspensions of actively moving particles. Cytoskeletal filaments interacting with a large number of surface-bound molecular motors give rise to dynamic bundle formation and stable swirling patterns[185,186]. In dense suspensions of motile bacteria, chaotic low-Reynolds-number flows have been observed on a mesoscopic scale, a phenomenon termed 'bacterial turbulence'[188,189]. Dense suspensions of swimming sperm cells at a glass-water interface can self-organize into vortex arrays with hexagonal order (spatial order), where additionally the sperm cells in each vortex phase-lock their flagellar oscillations (temporal order)[163]. The study of dynamic pattern formation in ensembles of active colloids and groups of organisms such as fish schools or bird swarms represents a sub-field of its own[190–193].

**Pattern formation in reaction-diffusion systems far from equilibrium.** A classical mechanism for the spontaneous formation of spatially inhomogeneous patterns are chemical reactions of diffusible molecules in spatially extended domains. This pattern formation requires a closed feedback loop between at least two reaction partners with different diffusion coefficients. Alan Turing proposed spontaneous pattern formation by reaction-diffusion-dynamics as a generic mechanism for the establishment of spatial patterns during the morphogenesis of organisms[40]. Recent experiments indeed revealed Turing mechanisms in a number of developmental processes[18], including pattern formation in the bacterial cytoskeleton[194], and the formation of digits during development[195], or the formation of stripe patterns in zebrafish[196,197].

The generic pattern formation mechanism of Turing patterns can be paraphrased as a principle of local activation and long-ranged inhibition[41]. A positive feedback amplifies the local concentration of an activator, where the activator concentration exceeds a certain threshold set by the concentration of an inhibitor. Fast diffusion of this inhibitor results in a long-ranged inhibitory effect that sets the size of activation regions. This mechanism can account for both stationary and dynamic patterns. A well-studied *in vitro* realization of this principle is the Beluzov-Zhabotinsky reaction, where cross-reacting and diffusing chemical species give rise to travelling wave patterns and rotating spirals in the presence of topological defects[198]. Obviously, the Beluzov-Zhabotinsky reaction represents a non-equilibrium system. Generalization of this mechanism, where spatial fields of active stresses, strains, or fluxes play the role of either activator or inhibitor had been already contemplated by Turing[40], with specific realizations proposed recently[42,43].

**Inside cells: self-assembly of cytoskeletal patterns.** Non-equilibrium dynamics in the cytoskeleton gives rise to the self-assembly of functional structures such as stress fibers or myofibrils on cellular scales. Below, we review selected examples of cytoskeletal pattern formation by local interactions inside cells. We will put special focus on myofibrillogenesis, *i.e.* the assembly of the almost crystalline acto-myosin bundles with sarcomeric periodicity in striated muscle cells.

*Regular patterns of the actin cytoskeleton.* Actin filaments can spontaneously form spatial patterns, both *in vitro* and inside living cells. Reconstituted actin filaments form stable bundles and asters as a result of passive depletion and active motor forces as well as entropic effects[199–202]. The interaction of actin filaments, crosslinkers, and myosin molecular motors results in dynamic patterns, including pulsatile myosin foci[203] and stable swirling patterns[185], revealing the rich dynamics of cytoskeletal pattern formation far from equilibrium.

Inside cells, actin filaments, actin-binding-proteins and myosin motors self-assemble into functional structures. A crosslinked meshwork of actin filaments with gel-like properties fills the intracellular space in animal cells and defines its rheological properties[204]. Myosin motors interact with actin filaments in a polarity-specific manner and exert microscopic force dipoles. As a result, myosin activity confers active contractility to the actin meshwork. We note that acto-myosin contractility is an emergent property of ensembles of interacting actin filaments and bipolar myosin motor filaments. A single myosin may either contract or expand a pair of parallel aligned actin filaments, depending on two possible configurations of structural polarity[205,206]. Specific physical mechanisms that effectively break the symmetry between expansion and contraction have been proposed, which result in a net compressive effect. These mechanisms include prolonged residence of myosin at the plus-end of actin filaments, filament rotation, and buckling of actin filaments under compressive load[205–207]. A thin and dense acto-myosin-meshwork beneath the cell membrane of animal cells constitutes the actin cortex. This actin cortex sets an effective surface tension and represents a major determinant of cell shape.

In cells that mechanically interact with an elastic substrate, actin filaments and myosin motor filaments form dense bundles, termed stress fibers[21]. These stress fibers generate contractile forces on a length-scale of cellular dimensions. Contractile actin-bundles also form the cell division ring that constricts animal cells during cytokinesis, the final stage of cell division. Contractile actin bundles can even span across multiple cells and exert contractile forces on the tissue scale[208]. In the stress fibers of certain cell types, such as fibroblast connective tissue cells, actin-crosslinkers and myosin motors are not distributed homogeneously along the fiber, but display periodic patterns with alternating localization with a characteristic periodicity of $1-2\,\mu$m [209]. The periodic patterns in these striated stress fibers are reminiscent of the sarcomeric arrangement of actin-crosslinkers and myosin in myofibrils, to be discussed in more detail below. We speculate that similar physical mechanisms of self-organized pattern formation account for the self-assembly of periodic patterns both in striated stress fibers and myofibrils[10]. Maturation processes that regularize initial periodic patterns may be lacking in striated stress fibers.

In summary, self-organized pattern formation of the actin cytoskeleton results in a diverse set of functional structures that are characterized by different types of spatial order. These include (figure 5)
- isotropic symmetry, *e.g.* in crosslinked actin meshworks and the actin cortex[202]
- nematic order, *e.g.* in stress fibers[21]
- smectic order, *e.g.* in striated stress fibers and myofibrils with sarcomeric periodicity[22].

*Actin binding proteins.* Actin filaments interact with a large number of actin-binding-proteins. Capping proteins regulate polymerization dynamics at the structural plus- and minus-end in a differential manner, and thus control filament length and treadmilling dynamics. Severing proteins such as gelsolin bind along the length of actin filaments and sever filaments at the binding position. Since their net rate of binding depends on filament length, severing represents a simple, yet effective mechanism of filament length control[210]. Crosslinking proteins promote the formation of crosslinked meshworks of actin filaments and the formation of actin bundles with nematically aligned filaments. Inside myofibrils, structural proteins such as tropomyosin decorate actin filaments for stability. Additional proteins provide structural support and elastic linkage inside sarcomeres. This includes the 'giant proteins' titin and nebulin, which scan span 0.5 $\mu$m in their extended configuration[211].

*Regular patterns of microtubules.* Microtubules interact with the actin cytoskeleton and have been proposed to represent tension-bearing structural elements[212]. During cell division, microtubules self-assemble the mitotic spindle[213], which constitutes part of the cell division machinery that distributes the chromosomes to the prospective daughter cells. This self-assembly process is driven by active motor forces and continuous turn-over of polymerizing and depolymerizing microtubules. Spindle assembly is orchestrated by two microtubule nucleation centers, which are located at opposite poles of the spindle. Each of these microtubule nucleation centers usually contains a centriole, a regular structure of triplet microtubules that the bears the same nine-fold symmetry as the microtubule-based axoneme of the eukaryotic flagellum. In fact, centrioles also serve as templates for axoneme assembly. Many cells, including the green alga *Chlamydomonas*, possess exactly two centrioles, which are used to assemble either a mitotic spindle during cell division, or to assemble and anchor up to two flagella. This shared use of centrioles implies that cell division and flagellar motility cannot occur at the same time in these cells[214]. The shared use of centrioles also point at a common evolutionary origin of the mitotic spindle and the flagellar axoneme[87].

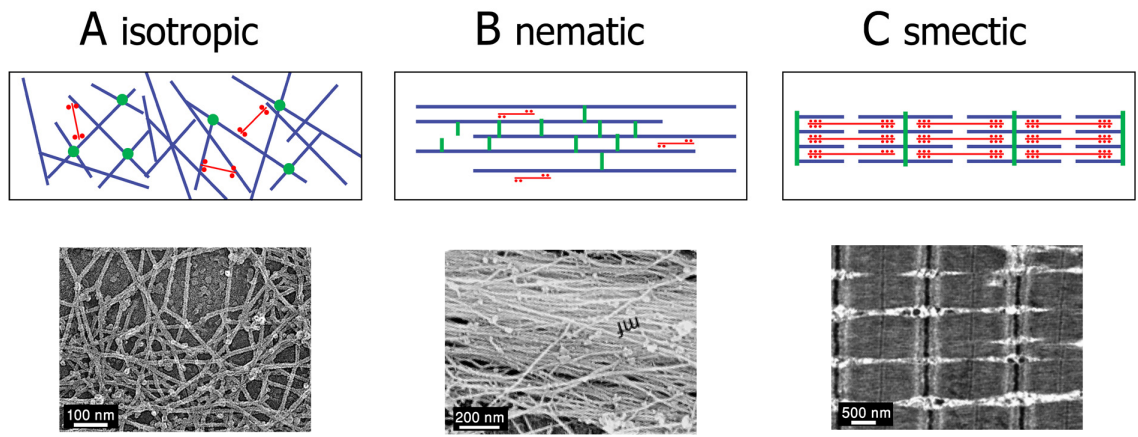

**Figure 5:** *Pattern formation in the actin cytoskeleton.* **A.** Actin filaments form dense crosslinked meshworks with isotropic symmetry, *e.g.* in lamellipodia or the actin cortex of animal cells. **B.** Actin filaments can organize into bundles of aligned filaments, representing a case of nematic order, *e.g.* in connective tissue cells. **C.** Inside striated muscle cells and cardiomyocytes, actin and myosin filaments are arranged in myofibrils of almost crystalline regularity. Myofibrils are characterized by periodic patterns of sarcomere units and represent a case of smectic order in the cytoskeleton. Micrographs from ref.[215,216] with permission.

**Example of cytoskeletal pattern formation: Self-assembly of myofibrils.** Myofibrils are the active force generator inside striated muscle cells and the cardiomyocytes of the heart[3]. Upon activation by calcium signals, myofibrils actively contract by the concerted activity of millions of myosin molecular motors inside. Myofibril contractions underlie all voluntary movements of higher animals and the rhythmic beat of the heart. The *de novo* assembly of myofibrils serves as a model system to study the general question how large scale patterns can emerge from local interactions in non-equilibrium systems.

Myofibrils are highly regular macro-molecular structures that are characterized by a periodic repetition of unit cells termed sarcomeres, see Figure 5-C. Sarcomeres are composed of actin filaments, myosin motor filaments, and additional structural proteins. Their regular spatial organization inside myofibrils represents a case of cytoskeletal order with almost crystalline regularity. Myofibrils measure $100-1000\,\mu$m in length; their functional and structural unit, the sarcomere, displays typical lengths of $1-2\,\mu$m. The major constituents of the sarcomere are actin and myosin filaments, the latter being polymerized out of individual muscle-specific myosin molecular motors. The structural plus ends of the actin filaments are anchored in a crosslinking region termed Z-band, with the actin-binding protein $\alpha$-actinin as an important constituent. During myofibril contraction, myosin filaments slide relative to the actin myosin filaments towards the actin plus-end, resulting in shortening of each sarcomere. This generation of active forces relies on the defined structural polarity of filaments inside each sarcomere. In addition to the three proteins named, actin, myosin, and $\alpha$-actinin, several hundred different proteins ensure the structural integrity and regulation of sarcomeric force generation. Interestingly, some of the largest known proteins are found inside myofibrils, such as the giant protein titin that spans half a sarcomeres length and serves as an elastic element[211]. Maximal force generation by myofibrils requires a dense and regular packing of myosin molecular motors and their actin tracks. In fact, the arrangement of proteins inside myofibrils are crystalline and result in distinct X-ray diffraction patterns[217].

It is an open question, how myofibrils are assembled. It has been proposed that existing myofibrils can grow by an epitaxy-like mechanism or serve as templates for the assembly of additional myofibrils. Additionally, precursor structures named premyofibrils, which already have a periodic structure, may be assembled first. These premyofibrils can then serve as template for mature myofibrils[218,219]. There is partial experimental evidence for the premyofibril hypothesis in at least some cell types. Yet, the fundamental question remains open: How do micrometer-sized building blocks such as actin filaments, myosin filaments, and titin assemble periodic structures, whether these are striated stress fibers, premyofibrils, or nascent myofibrils? It can be considered certain that giant scaffolding proteins such as titin and obscurin serve as a molecular templates for the assembly and structural organization of single sarcomeres[211,220]. However, it is unclear if these scaffolding proteins are involved already in the early establishment of periodic patterns in initially unstriated acto-myosin bundles. The proposition that that a periodic arrangement of titin molecules directs myofibrillogenesis would require a yet unknown mechanism by which titin molecules become arranged in periodic patterns first. We emphasize that the diffusion coefficients of large molecules such as titin or actin and myosin filaments are extremely low, which renders their passive sorting into periodic patterns kinetically impossible. We thus argue that active, force-generating processes should be required for myofibrillogenesis. This hypothesis is consistent with recent experimental findings that emphasize the importance of active tension for myofibrillar pattern formation, and the requirement for attachment to support structures, such as tendons, which can resist active forces[221].

Myosin motor forces are an obvious source of active force generation inside nascent myofibrils. Yet, these forces cannot explain self-assembly. In myofibrils, myosin filaments are localized near the structural minus-end of actin filaments, despite their tendency to slide towards the plus-end. We speculate that strong myosin forces can disrupt nascent myofibrils. Interestingly, it has indeed been found that myosin force generation is up-regulated only after the initial stages of myofibril assembly. In some cells, myofibril precursors are assembled with non-muscle myosin filaments that generate less force, which are replaced by muscle-myosin only at later stages. These findings are consistent with a potentially destructive role of myosin forces during myofibril assembly. Different physical mechanisms have been proposed for the source of forces that sort actin and myosin filaments in place in nascent myofibrils. Zemel *et al.* proposed a sorting mechanism that depends on a second, hypothetical motor, which is minus-end directed[222]. This model could indeed account for the spontaneous emergence of periodic structures with polarity-sorted filaments. However, the involvement of such a hypothetical minus-end directed motor needs yet to be demonstrated. Yoshinaga *et al.* proposed a Turing-like mechanisms with a mutual coupling of fields of local active stress and actin polarity[42]. This generic mechanism predicts the emergence of periodic polarity patterns, yet its coarse-grained nature does not inform about underlying molecular mechanisms. In publication 3.1, we present a minimal model for the self-assembly of periodic cytoskeletal patterns as observed in myofibrils from local interactions between three constituents: actin filaments, bipolar myosin filaments, and a plus-end actin crosslinker[10].

**Inside tissues and organisms: self-organized morphogen gradients.** Inside tissues and organisms, concentration gradients of signaling molecules regulate growth and developmental patterning[223]. Important examples include the proteins Bmp[224], Dpp[225], and Wnt[224], which establish concentration gradients along principal body axes during embryonic development. Signaling molecules that spatially orchestrate cell differentiation during morphogenesis are termed morphogens, a term first coined as a theoretical concept by Turing[40]. A more biological view defines morphogens as signaling molecules that are secreted in localized source regions and form long-ranged concentration gradients that determine discrete cell fates in a concentration-dependent manner[223,226] (although this definition may disqualify some signaling molecules such as Wnt as classical morphogens[227]).

In a minimal description, a morphogen is secreted at a point source located at $x = 0$, and diffuses with effective diffusion coefficient $D$ in a spatial domain of size $L$, while being subject to degradation with degradation rate $k$

$$\dot{c} = p\delta(x) - kc + D\nabla^2 c. \tag{5}$$

Note the effective diffusion term can account also for undirected active transport, *e.g.* by cellular transport processes of endocytosis and exocytosis[228], in addition to passive diffusion. Equation (5) implies an exponential concentration profile $c(x) \sim \exp(-x/\lambda)$, where the pattern length-scale $\lambda$ is set by a competition of diffusion and degradation

$$\lambda = \sqrt{D/k}. \tag{6}$$

Remarkably, scaling of concentration profiles with system size $L$, characterized by $\lambda \sim L$, has been observed in a number of biological systems, including the developing fly wing[229–231]. In the fly wing, it was shown that pattern scaling results from a dynamic regulation the morphogen degradation rate according to system sizes, $k \sim L^{-2}$ [229]. Several theoretical mechanisms for self-regulated pattern scaling have been proposed[232–234], which still need to be experimentally confirmed.

These theoretical mechanisms of pattern scaling are challenged in systems with regeneration capabilities. For example, small amputation fragments of the flatworm *Schmidtea mediterranea* can regenerate into a miniature version of the original worm with a proportional body plan proportionally scaled according to the size of the amputation fragments[235]. This re-patterning occurs within weeks. Additionally, flatworms can scale their body plan by a factor of 20 in length during growth and active degrowth, depending on feeding conditions[235]. Long-range gradients of gene expression patterns of signaling molecules such as Wnt are known to pattern the anterior-posterior-axis in flatworms[236]. Regeneration capabilities as observed in flatworms require de-novo formation of a morphogen source after amputation. Mechanisms of self-organized pattern formation such as Turing mechanisms can account for the formation and positioning of a new morphogen source[40,41]. However, classical Turing mechanisms are characterized by fixed intrinsic pattern length-scales, which are again set by a competition of diffusion and degradation of patterning molecules, compare equation (6). In publication 3.2, we present a general mechanism for self-organized pattern formation that generates spatial patterns that scale with system size[11]. This minimal mechanism displays structural robustness and can cope with parameter variations and fluctuations. We review different sources of fluctuations in biological systems.

## 1.4 Fluctuations and biological robustness

Life relies on stochastic processes[237]. Thermal noise enables diffusive transport and biochemical reactions at the molecular scale. Small-number fluctuations cause stochastic dynamics at the scale of cells and organisms. Even biological evolution relies on fluctuations: stochastic events during gene duplication generate genotypic variations. Here, we will focus on cell motility, which is driven by stochastic non-equilibrium dynamics of its cytoskeleton. This implies measurable active fluctuations at the mesoscopic scale of the cell that violate the fluctuation-dissipation theorem. A common phenomenological description of active fluctuations in terms of an effective temperature $T_{\text{eff}}$ above the thermodynamic temperature $T$ can provide a first, rough approximation only. Effective descriptions of noise in biological systems, which coarse-grain chaotic out-of-equilibrium dynamics at the microscopic scale, are researched actively[13,14].

### 1.4.1 Sources of fluctuations in biological systems

Any chemical reaction between molecules requires that Brownian motion first brings the reaction partners into close contact. Thermal fluctuations are also required to overcome energetic barriers of the chemical reaction. In consequence, each reaction step is a stochastic event, which can often be described as a Poisson jump process[238]. Small-number fluctuations of chemical reactions do not necessarily average out at the scale of the cell. One simple reason for this is that copy numbers inside cells can be as low as a few tens or hundreds[239]. Thus, mean field description may miss important aspects of cellular dynamics. As an important aspect, nonlinear feedback loops can amplify small-number fluctuations, thereby propagating fluctuations from the molecular scale to the mesoscopic scale of the cell. For example, the stochastic binding of a single transcription factor to a specific DNA binding site can initiate the transcription of a particular gene and result in its translation into many copies of the corresponding protein. Such stochastic gene expression can lead to different gene expression profiles in cells of identical genetic setup[240]. Signaling systems that implement bistable switches with long hysteresis can amplify this effect. In fact, some bacterial colonies harness stochastic gene expression to induce phenotypic heterogeneity within the population, which can provide a competitive advantage in a time-varying environment[241]. In addition to intracellular noise, cells are subject to external perturbations. These external perturbations include fluctuations in nutrient levels and physical parameters such as light, pH, and temperature, each of which affects the dynamical state of the cell.

**Thermal fluctuations.** Free energy differences of biochemical reactions are commonly on the order of a few $k_B T$, where the thermal energy $k_B T \approx 4 \times 10^{-21} J$ at room temperature. For example, the Gibbs free energy difference for the hydrolysis of a single ATP molecule equals $20 - 25\ k_B T$ [63]. Unfavorable reactions inside cells that result in high-energy reaction products, a local reduction of entropy, or which perform mechanical work, are coupled to the hydrolysis of triphosphate nucleosides such as ATP, which breaks detailed balance of reaction cycles[3,24,63]. Such reactions include the chemo-mechanical cycles of molecular motors[242]. As a specific example, single kinesin molecular motors take directed 8 nm steps on a microtubule per chemo-mechanical cycle, which corresponds to mechanical work of $10\,k_B T$ at full 5 pN load force. Molecular motors that show on average directed motion will occasionally take a backward step due to thermal fluctuations.

**Molecular shot noise and small-number fluctuations.** Inside cells or subcellular compartments, copy numbers of proteins are often in the range of hundreds. For example, the volume of an *E. coli* bacterium comprises just a few femto-liters, which implies that a concentration of one micro-molar corresponds to just a few thousand molecules. This is a typical order of magnitude for the most abundant bacterial proteins. Eukaryotic cells can be much larger than bacteria, yet important signaling processes are often spatially confined to sub-cellular regions of femto-liter volume, such as the nucleus or a flagellum. Low copy numbers imply substantial small-number fluctuations of chemical reactions and thus introduce noise in cellular signaling. Inside tissues, communication between cells is subject to the same sources of noise inherent to chemical reactions[243].

**Sensory perception at the physical limit.** Fluctuations are paramount in sensory perception of weak stimuli. Many sensory organelles can operate at the physical limit[244]. For example, rod photoreceptors in the retina of the eye can detect single photons[245–247]. Hair bundles in the inner ear respond to faint vibrations that carry an energy of only a few $k_B T$ per oscillation cycle[244]. Specialized olfactory sensory neurons can detect single odorant molecules[248], likewise sperm respond to single chemoattractant molecules[249]. Such signaling events are inherently stochastic in nature.

In addition to this quantized nature of single molecule detection, thermal fluctuations impact on sensation at the physical limit: the absorption of a single photon or the binding of a single ligand molecule to a receptor induces a conformational change in the receptor proteins, which then activates down-stream signaling cascades. Thermal noise can induce the same conformational change as a detection event, and thus limits the precision of cellular signal detection[250,251].

**Active motor fluctuations.** Individual molecular motors progress through their mechano-chemical cycles in an inherently stochastic manner. Single-molecule experiments allow to detect discrete steps of single molecular motors and the stochastic timing of their stepping[252]. The collective dynamics in ensembles of molecular motors gives rise to active contractility, directed transport, but also to non-equilibrium fluctuations. A hallmark of non-equilibrium fluctuations is the violation of the fluctuation-dissipation theorem. At thermal equilibrium, the fluctuation-dissipation theorem relates the fluctuation spectrum $\langle |A(\omega)|^2 \rangle$ of a degree of freedom $A$ to its response function $\chi_A(\omega)$, which characterizes the response to an external perturbation[253]

$$\langle |A(\omega)|^2 \rangle = \frac{2k_B T}{\omega}\,\mathrm{Im}\,\chi_A(\omega). \tag{7}$$

Violations of the fluctuation-dissipation theorem have been experimentally observed in fluctuation spectrum of the cell membrane of red blood cells[254], fluctuations of the cytoskeleton[255–257], or the hair bundles of auditory sensory neurons in the inner ear[258]. These non-equilibrium fluctuations have been attributed to the stochastic collective dynamics of molecular motors.

The beat of the eukaryotic flagellum exhibits active fluctuations as well. Previous studies reported Fourier peaks of finite width in power spectra of flagellar dynamics, which provides a signature of phase fluctuations. Goldstein *et al.* conducted an indirect measurement of these phase fluctuations by examining the frequency of phase-slips in pairs of synchronized flagella[160,259,260]. In publication 2.3, we present a method to measure phase and amplitude fluctuations of the flagellar beat directly. Specifically, our method maps high-precision measurements of flagellar bending waves on a generic description of a limit cycle oscillator with phase and amplitude noise, the normal form of a Hopf bifurcation with complex noise term,

$$\dot{Z} = i(\omega_c - \omega_1 |Z|^2)Z + \mu(\Lambda - |Z|^2)Z + (\xi_A + i\xi_\varphi)Z. \tag{8}$$

Here, $Z = Ae^{i\varphi}$ is a complex oscillator variable and $\xi_A$ and $\xi_\varphi$ denote amplitude and phase noise terms, respectively. In publication 2.3, we further consider a minimal model of stochastic collective motor dynamics and show that emergent noisy motor oscillations can be likewise mapped on equation (8) of a noisy Hopf oscillator.

Stochastic motor dynamics has been studied in a number of systems, ranging from stochastic oscillations of hair bundles of auditory sensory neurons in the inner ear[34], bi-directional transport of actin filaments that interact with a surface coated with myosin motors in motility assays[261,262], to oscillations in in vitro system of myosin motors and actin filaments[263] and 'artificial cilia' consisting of microtubule bundles interacting with kinesin motors[187]. These measurements of non-equilibrium fluctuations in mesoscopic systems provide a way to observe signatures of the stochastic dynamics of molecular motors.

### 1.4.2 Example of stochastic dynamics: synchronization of noisy oscillators

Since the observation of phase-locked pendulum clocks by van Huygens, it is known that active oscillators can synchronize by virtue of a weak coupling, despite effects of noise or a mismatch of intrinsic oscillation frequencies[264]. In the middle of the 20th century, work on generators of radio signals, and phase-locked electronic oscillators in particular, motivated the development of a general theory of synchronization[265,266]. Synchronization is also observed in biological systems: examples include synchronization of the walking gaits of pedestrians[268], coupling of the 'segmentation clock' genetic oscillators that orchestrate somatogenesis during embryonic development[267], and last but not least synchronization in collections of beating flagella as studied in this thesis. Synchronization of oscillators implies the emergence of a common oscillation frequency and a fixed phase relation. We first review the Adler equation, a generic description for the synchronization of a pair of coupled noisy oscillators in the next paragraph and then turn to the synchronization in pairs of beating flagella in the remainder of this section.

**The stochastic Adler equation of coupled oscillators.** The stochastic Adler equation provides a generic description for the synchronization of a pair of noisy, active oscillators. The dynamics of the phase difference $\delta = \varphi_1 - \varphi_2$ between the two phase oscillators with respective phases $\varphi_1$ and $\varphi_2$ can be idealized by[265]

$$\dot{\delta} = \Delta\omega - \lambda \sin\delta + \xi. \tag{9}$$

Here, $\Delta\omega = \omega_1 - \omega_2$ denotes the mismatch between the intrinsic frequencies of the two oscillators, $\lambda$ is an effective coupling strength and $\xi(t)$ denotes a Gaussian white noise term with $\langle\xi(t)\xi(t')\rangle = 2D\,\delta(t-t')$. The Adler equation, equation (9), captures key dynamical features of synchronization, which we discuss below. Many specific systems of coupled oscillators can be approximately mapped on the Adler equation. This includes a description of flagellar swimming and flagellar synchronization in free-swimming *Chlamydomonas* cells presented in publication 2.2. The dynamics given by equation (9) can be interpreted as that of an overdamped particle in a tilted periodic potential $U(\delta) = -\Delta\omega\delta - \lambda\cos\delta$ [269]. For zero frequency mismatch, $\Delta\omega = 0$, steady states of equation (9) correspond to in-phase synchronization with $\delta = 0$, and anti-phase synchronization with $\delta = \pi$. For positive synchronization strength $\lambda > 0$, in-phase synchronization is stable, while anti-phase synchronization represents a meta-stable steady state, see figure 6-A,B. Noise induces stochastic phase slips at a frequency $G = D/|2\pi I_0(\lambda/D)|^2$ [266], where $I_0$ denotes the modified Bessel function of the first kind. In case of a frequency mismatch $\Delta\omega \neq 0$, the two oscillators will synchronize with a phase-lag $\delta^* = \sin^{-1}(\Delta\omega/\lambda)$ at steady-state, provided $|\Delta\omega| < |\lambda|$. If the frequency mismatch becomes too large, the system undergoes a saddle-node bifurcation and no steady state exists anymore for $|\Delta\omega| > |\lambda|$. In this case, the dynamics is characterized by phase drift, see figure 6-D. Many analytic results for the stochastic Adler equation are known, see *e.g.* the book by Stratonovich[266].

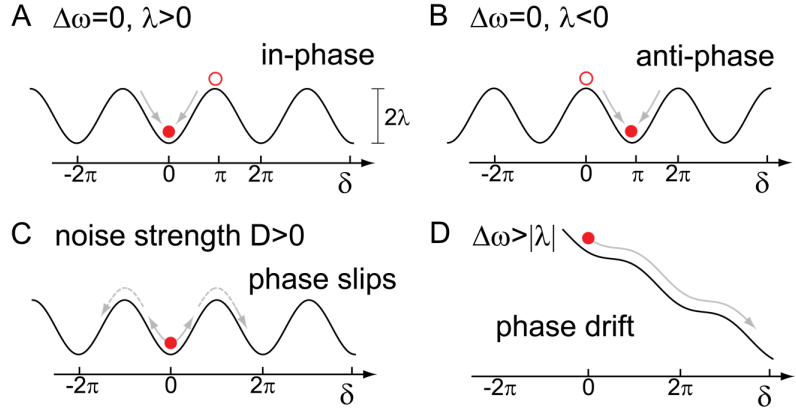

**Figure 6:** *Dynamic regimes of the stochastic Adler equation.* The dynamics of the phase difference $\delta$ is equivalent to the overdamped motion of a particle in a tilted periodic potential. **A.** For positive coupling strength $\lambda > 0$, the in-phase synchronized state is stable, while the state of anti-phase synchronization is meta-stable. **B.** For negative synchronization strength $\lambda < 0$, stability is reversed. **C.** In the presence of noise, d exhibits stochastic transitions between adjacent stable states, so-called phase slips. **D.** If the mismatch $\Delta\omega$ between the intrinsic frequencies of the two coupled oscillators becomes too large, no synchronization occurs, and the phase difference will increase monotonically, corresponding to a regime of phase drift. Modified from[273].

*Flagellar synchronization and the Adler equation.* All dynamic regimes predicted by the Adler equation have been observed for pairs of beating flagella, including in-phase[159,160,270] and anti-phase synchronization[271], stable phase-lags[160,161], phase slips[159,160,259] and phase drift[5,272]. Synchronization of beating flagella have been proposed to result from a mechanical coupling between flagella, *e.g.* by direct hydrodynamic interactions[4]. The symmetry of the Stokes equation, equation (3), which governs hydrodynamics at the cellular scale, prompts systems that explicitly break either time-reversal symmetry or spatial symmetries to facilitate such hydrodynamic synchronization.

**Hydrodynamic synchronization requires broken symmetries.** Already in 1951, Taylor proposed that the remarkable phenomenon of flagellar synchronization arises from a mechanical coupling between flagella, such as direct hydrodynamic interactions[4]. Only recently was flagellar synchronization by direct hydrodynamic interactions unequivocally demonstrated in experiments. Using pairs of flagellated cells held in separate micro-pipettes at a distance, it was found that flagellar synchronized with a distance-dependent synchronization strength[6]. Synchronization by direct hydrodynamic interactions was additionally studied in systems of artificial actuators such as colloids driven by oscillating magnetic fields, or 'light-mill' micro-rotors driven by laser-light[274–278].

Similar to the problem of self-propulsion at low Reynolds numbers, hydrodynamic synchronization requires broken symmetries to overcome the symmetries of the Stokes equation and to provide a net coupling between oscillators. For illustration, we consider the idealized example of two spheres revolving around circular trajectories. Each sphere is driven by a constant tangential force and would thus assume a constant angular speed $d\varphi/dt = \omega_0$ if the other sphere were absent. The motion of one sphere generates a long-ranged flow field whose strength decays with inverse distance. This flow field exerts a hydrodynamic interaction force on the second sphere, which changes the phase speed $d\varphi/dt$ of this sphere. Although, these hydrodynamic interactions indeed couple the phase dynamics of the two spheres, the resultant net coupling strength $\lambda$ is zero. This can be seen from symmetry arguments[122]: a spatial mirror operation and time-reversal will both result in the same dynamics, since the Stokes equation is invariant under these operations. While the spatial mirror operation does not change the stability of any synchronized state, stability is reversed under time-reversal. We conclude that any synchronized state can be neither stable nor unstable, hence $\lambda = 0$.

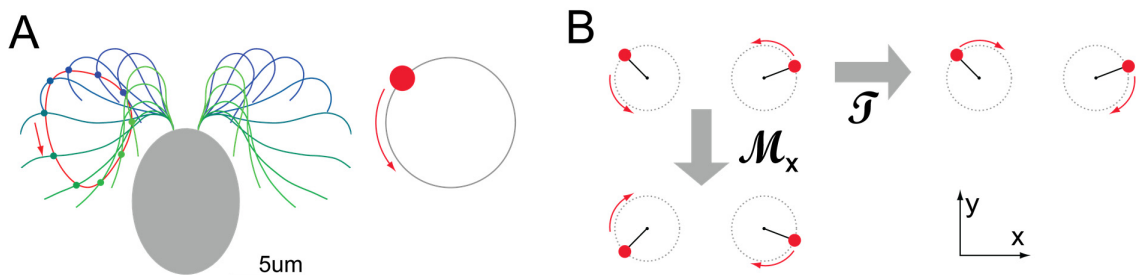

**Figure 7:** *Lack of hydrodynamic synchronization in a minimal model with symmetries.* **A.** In an idealized model, a beating flagellum is represented by a sphere that moves in a viscous fluid along a circle, being driven by a constant tangential force [112]. This model is inspired by the observation that each point on a beating flagellum moves on a circular orbit. **B.** Direct hydrodynamic interactions between the two rotating spheres couple their motion. However, the net synchronization effect is zero, as a consequence of symmetries: A reflection $\mathcal{M}_x$ of the system at the $x$-axis is dynamically equivalent to a time-reversal $\mathcal{T}$. As the time-reversal reverses the stability of synchronized states, while a reflection does not, we conclude that any synchronized states is neutrally stable[112,115,122].

Different physical mechanisms have been proposed for synchronization by direct hydrodynamic interactions, all of which break spatio-temporal symmetry in one way or the other, which we review now.

*Amplitude compliance.* In a generalization of the two sphere model considered by Lenz *et al.*, the radii of the circular tracks are not constant, but are considered as elastic degrees of freedom with an effective stiffness $k$ [279]. This amplitude compliance breaks the time-reversal symmetry of the equation of motion. It is found that both spheres can synchronize their motion with an effective synchronization strength that scales inversely with the amplitude stiffness, $\lambda \sim 1/k$.

*Phase-dependent driving forces.* Golestanian *et al.* proposed a different symmetry-breaking mechanism that relies on phase-dependent driving forces[280,281]. Thereby, the two sphere system is not invariant anymore under a spatial mirror operation, thus facilitating net synchronization.

*Other means to break symmetry.* Additionally, Theers *et al.*, considered the effect of small, non-zero Reynolds number[282]. The effect of unsteady acceleration at finite Reynolds numbers breaks time-reversal symmetry and results in net synchronization with a synchronization strength $\lambda \sim \text{Re}^{1/2}$. In the original formulation of the two sphere model by Vilfan *et al.*, a no-slip boundary close to the two spheres was introduced, which breaks spatial mirror symmetry[112].

The beat of real cilia and flagella is characterized by both phase-dependent driving forces and a finite compliance of the flagellar wave forms[118,151]. Thus, time-reversal symmetry is broken, which allows for flagellar synchronization by direct hydrodynamic interactions. It should be noted that the symmetry breaking can be weak, thus resulting in a weak synchronization strength[5,156]. In publication 2.1, we present an alternative synchronization mechanism that operates independently of direct hydrodynamic interactions[283]. Instead, this mechanism relies on a closed feedback loop between flagellar dynamics and swimming motion. In publication 2.2, we demonstrate that this synchronization mechanism is important in free-swimming *Chlamydomonas* cells[5]. Synchronized beating of its two flagella is a prerequisite for this cell to swim fast and straight.

Purposeful motion further requires a control of swimming direction in response to environmental cues. In the next section, we review navigation mechanisms for directed motion in external concentration gradients.

### 1.4.3 Cellular navigation strategies reveal adaptation to noise

We review three distinct strategies employed by single cells for navigation in external concentration gradients[8]. Any cellular gradient-sensing strategies must cope with noise, such as motility fluctuations, or molecular shot noise of chemosensation at dilute concentrations. We argue that the three different chemotaxis strategies employed by single cells represent an adaptation to the respective strength of motility and sensing noise encountered by these cells.

**Physical limits to chemo-sensation.** Cells constantly monitor extracellular concentrations of signaling molecules. This sensory input controls *e.g.* chemotaxis in chemical gradients or cell fate decisions during development. Berg and Purcell made seminal contributions to our understanding of the physical limits of chemosensation[284]. Their work was later re-derived in the framework of statistical physics by Bialek *et al.*[250]. Specifically, many cells measure the extracellular concentration $c$ of a signaling molecule by counting binding events of individual molecules to cognate surface receptors. In the idealized limit of maximal uptake, characterized by high receptor density and irreversible binding, the mean number $\langle N \rangle$ of binding events in a time window of duration $\tau$ is given by[250,284]

$$\langle N \rangle = 2\pi L \cdot Dc \cdot \tau, \tag{10}$$

where $D$ denotes the diffusion coefficient of the molecule. The geometric factor $2\pi L$ corresponds to a cellular geometry of a perfect sphere of diameter $L$; generally it will scale with the longest dimension of the cell[285]. Individual binding events will be uncorrelated to good approximation, and thus the actual number $N$ of binding events will be a Poissonian random variable with mean $\langle N \rangle$ and variance $\langle N \rangle$. Cellular concentration measurements are thus prone to molecular shot noise: at physiological pico-molar concentrations, only a few molecules per second will diffuse to the cell. Hence, cells face trade-off choices between the accuracy of concentration measurements and the temporal resolution for sensing time-varying concentrations, which has implications for cellular chemotaxis strategies.

For chemotaxis, the maximal integration time $\tau$ for local concentration measurements is set by the time it takes the cell to move one body length, $\tau \sim L/v$, where $v$ and $L$ denote speed and size of the cell. As a numeric example, we find $N \approx 500$ from equation (10) using typical values for a bacterial cell ( $L \approx 3\,\mu\text{m}$, $L \approx 50\,\mu\text{m}\,\text{s}^{-1}$, $D \approx 700\,\mu\text{m}^2\,\text{s}^{-1}$, $c = 1\,\text{nM}$). Thus, concentration measurements are inherently unreliable at the cellular scale. This constrains the range of potential chemotaxis strategies for a given cell.

*Motility noise.* Motility noise comprises contributions from both thermal fluctuations and active motility fluctuations. For the smallest cells, such as bacteria, the contribution from thermal fluctuations can be substantial. In particular, effective rotational diffusion will randomize the swimming direction of a cell. We can estimate the effect of thermal fluctuations on a passive particle of same shape as the cell, which provides a lower bound for the effective rotational diffusion coefficient. For such a passive particle, the rotational diffusion coefficient is given by an Einstein-Stokes relation, $D_{\text{rot}} = k_B T / \gamma_{\text{rot}}$, where $\gamma_{\text{rot}}$ denotes the hydrodynamic friction coefficient for rotational motion. The rotational friction coefficient typically scales as $\gamma_{\text{rot}} \sim L^3$, where $L$ denotes a typical size of the cell. Hence,

$$D_{\text{rot}} \sim L^{-3}. \tag{11}$$

As a numerical example, we find $D_{\text{rot}} = 0.1\,\text{s}^{-1}$, for a spherical particle of radius $a = 1$, for which $\gamma_{\text{rot}} = 8\pi\eta a^3$. This estimate implies that the correlation time $\tau = 1/D_{\text{rot}}$ of persistent directional swimming is just a few seconds for a micron-sized bacterium. For typical swimming speeds of a bacterium, $v \approx 10\,\mu\text{m}\,\text{s}^{-1}$, the resultant swimming path will be a persistent random walk (even in the absence of active tumbling) with persistence length of $l_p = v/(2D_{\text{rot}}) \approx 50\,\mu\text{m}$ [286,287]. It was argued that active motility would not pay off for the smallest bacteria, which measure less than a micron in size[288]: for these cells, directional persistence of motion would be so low that net locomotion becomes impossible and active motility would increase only the effective translational diffusion coefficient of the cell, $D_{\text{eff}} = D_{\text{trans}} + v^2/(6D_{\text{rot}})$. Interestingly, most cells that measure less than a micron are indeed immotile[288].

Motile cells, such as swimming bacteria, often show chemotaxis, *i.e.* the directed motility upwards a concentration gradient, such as a gradient of nutrient concentration. Rotational diffusion restricts the gradient sensing strategies available to these cells. During a time span on the order of $\tau$, the information gathered by swimming bacteria such as *E. coli* is not sufficient for directed steering responses in the direction of the gradient. Instead, these bacteria employ a stochastic navigation strategy, where only the frequency of random re-orientation events is adjusted. This steering strategy results in a biased random walk with net drift up the gradient as detailed in the next section.

**Three cellular strategies of gradient sensing.** Motility control of motile cells is an ideal test case to study the adaptation of cellular signaling to dynamic and noisy environments. Pioneering work by Howard Berg unraveled the stochastic control logic of chemotaxis in bacteria upwards concentration gradients of nutrients[81]. Motile bacteria such as *E. coli* perform a biased random walk to move up a chemical gradient. We enjoy a rather comprehensive understanding of chemotactic signaling in bacteria today[289,290]. Eukaryotic (*i.e.* non-bacterial) cells, however, employ fundamentally distinct navigation strategies of helical chemotaxis[140,291] and spatial comparison[61,292], which are reviewed below. For eukaryotic chemotaxis, many questions regarding the underlying sensorimotor feedback logic, and its adaptation to dynamic chemoattractant fields are still open. Below, we elaborate the hypothesis that different chemotaxis strategies of bacteria and eukaryotic cells actually represent adaptations to different regimes of noise in sensing and motility[8]. We emphasize that molecular shot noise makes renders measuring a concentration gradient accurately a non-trivial task at the cellular scale.

Measuring a concentration gradient requires the comparison of local concentration measurements $c_1 = c(\mathbf{r}_1, t_1)$ and $c_2 = c(\mathbf{r}_2, t_2)$ at different positions in space and possibly different times. The most direct method of gradient-sensing would be to measure local concentrations at different positions $\mathbf{r}_1$ and $\mathbf{r}_2$ of the cell at the same time $t_1 = t_2$, which amounts to a mechanism of *spatial comparison*. However, recalling that cells detect a number $N$ of binding events as a proxy for local concentration $c$ with $\langle N \rangle \sim c$, see equation (10), we are led to compare the difference in the expectation values of two measurements $\Delta N = \langle N_1 - N_2 \rangle$, to the uncertainty $\sigma$ in its measurement, which provides a signal-to-noise ratio[284]

$$\frac{\Delta N}{\sigma} \sim L^2 \sqrt{\frac{Dc}{v}} \frac{\nabla c}{c}. \qquad (12)$$

From this equation, we find that spatial comparison is a viable strategy only for relatively large and slowly moving cells. For fast-swimming bacteria and sperm cells, one finds that the signal-to-noise ratio of spatial comparison would be too low to allow for reliable steering. Instead, these cells employ different strategies of *temporal comparison*[8,293], for which these cells rely on their active motion inside the external concentration field, which allows them to compare concentrations at different positions along their swimming path $\mathbf{r}(t)$.

Noise in sensing and motility imposes tight constraints on cellular chemotaxis strategies. We review three different strategies for dynamic gradient sensing employed by single cells.

**Chemotaxis by spatial comparison.** The fidelity of spatial gradient sensing across the diameter $L$ of a cell depends strongly on the time $\tau = L/v$ available to integrate noisy local concentration measurements, which in turn depends on the speed $v$ of the cell locomotion[284]. Only slow moving cells, such as the slime mold *Dictyostelium* ($v \approx 1-10\,\mu\mathrm{m}/\mathrm{min}$), are able to employ spatial comparison for directed motion up a chemical gradient[61] with the efficacy of chemotaxis depending on the signal-to-noise ratio of spatial gradient sensing[294,295].

**Biased random walks.** Swimming bacteria such as *Escherichia coli* employ a stochastic chemotaxis strategy: they move along biased random walks to steer up chemical gradients, *e.g.* gradients of nutrient concentrations. During so-called 'run' periods, these cells swim along straight paths for a few seconds. These straight 'runs' are interrupted by stochastic reorientation events, termed 'tumbling', during which the bacterium picks a new swimming direction at random[69]. *E. coli* employs a particular form of temporal comparison for gradient-sensing by which the cell computes a smoothed time-derivative of the temporal concentration signal $c(\mathbf{r}(t))$ along its swimming path $\mathbf{r}(t)$. If a decrease of this concentration signal is detected, which is indicative of inadvertently swimming down-gradient, the cell will tumble earlier and more vigorously. This 'run-&-tumble' strategy results in a biased random walk, with net drift towards regions of higher chemoattractant concentration.

Interestingly, noise in chemosensation would render the alternative chemotaxis strategy of spatial comparison too inaccurate for these bacteria[69,284], see also equation (12). In short, *E. coli* is too small and too fast for accurate gradient-sensing by spatial comparison. At the same time, motility fluctuations are similarly substantial for these cells: Sized only a few microns, *E. coli* cells are subject to thermal fluctuations that randomize their swimming direction even during supposedly straight 'runs'. The rotational diffusion time $D_{\mathrm{rot}}^{-1}$ of a few seconds limits the available time span for signal integration. Correspondingly, it is observed that 'runs' usually do not last longer than this time span. The information gathered by temporal comparison during a single 'run' is not sufficient to control the steering direction[284], which leads to the genuinely stochastic strategy of bacterial chemotaxis. Thus, sensing and motility fluctuations constrain the possible choice of chemotaxis strategy for bacteria.

**Helical chemotaxis.** A third navigation strategy exploits chiral self-motion. This strategy is employed *e.g.* by sperm from marine species with external fertilization, which respond to signaling molecules released by the egg[144,296]. These sperm swim along helical paths[91–93], which is a result of the chiral beat of their flagellum[93]. Helical swimming enables these cells to detect the direction of a chemoattractant gradient perpendicular to the helix axis: when the cell swims along a helix (whose axis is initially not aligned with the gradient), the cell will periodically move up and down the gradient, see figure 9. Thus, the cell perceives a chemoattractant stimulus that oscillates with the frequency of helical swimming. This frequency is about 2 Hz for marine sperm[119]. Thereby, information about the spatial gradient becomes encoded in a temporal oscillation of the chemoattractant signal. This chemoattractant signal is transduced by a chemotactic signaling cascade[297], which generically will elicit an oscillatory flagellar steering response[140]. As a result, the curvature and torsion of the sperm swimming path oscillate with the helix frequency. While a constant value of curvature and torsion characterizes a perfect helix, oscillations of curvature or torsion result in bending helices[140]. As a consequence, the helix axis, which represents the direction of net motion, aligns with the gradient direction. Correct steering requires that the latency time of chemotactic signaling, which induces a phase-shift between stimulus oscillations and curvature oscillations, adopts an optimal value[140].

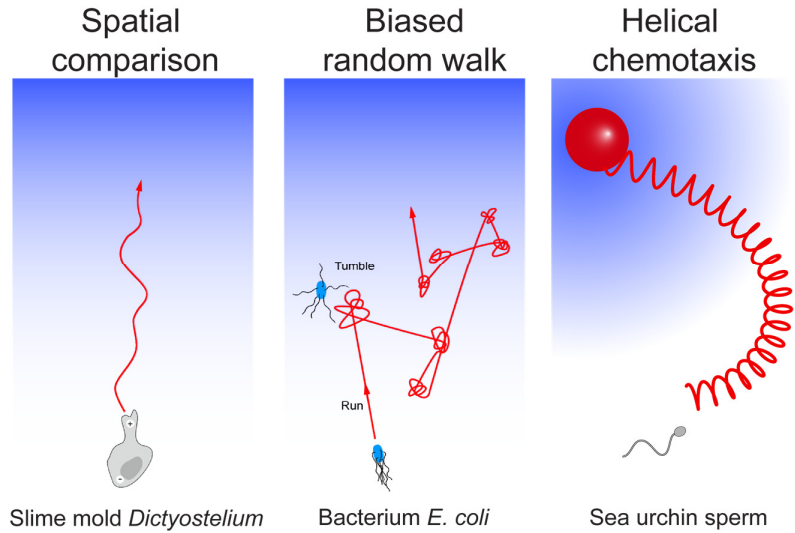

**Figure 8:** *The three strategies of single cell chemotaxis.* For chemotaxis, motile cells employ very different strategies. *Left*: Cells with crawling motility, such as the slime mold *Dictyostelium,* compares concentration differences across the diameter of the cell, thus representing a case of *spatial comparison.* Subsequently, the cell becomes structurally polar in the direction of the gradient and moves up-gradient. *Middle:* The bacterium *E. coli* employs a chemotaxis strategy of temporal comparison along a *biased random walk*. Short run segments are interrupted by stochastic reorientations events termed 'tumble'. A dynamic regulation of the 'tumbling' frequency according to temporal changes of the concentration signal along their swimming path results in a net drift up the gradient. *Right:* Sperm cells from marine species with external fertilization navigate along helical paths, which represents a stereotypic form of exploratory motion. Although *helical chemotaxis* represents a case of gradient-sensing by temporal comparison as in the case of bacterial chemotaxis, sperm steering responses are deterministic and point in the direction of the gradient, in contrast to the stochastic tumbling events of bacteria. The different chemotaxis strategies of these three cells suggest an adaptation to different noise levels of sensing and motility. Blue: concentration gradient of a signaling molecule, red: cell trajectory. Modified from ref.[8] with permission.

Sperm cells swim too fast ($v = 100\,\mu\text{m/s}$) in order to employ spatial comparison with sufficient accuracy along the length of their flagellum ($L = 50\,\mu\text{m}$)[8]. Like bacteria, sperm must rely on temporal comparison, *i.e.* sperm detect how the local concentration changes in time while they actively move in a concentration gradient[298]. However, being ten-fold larger than bacteria, sperm cells are 1000-times less affected by thermal rotational diffusion[8]. Thus, sperm cells from marine invertebrates can stably swim along helical paths. Their helical swimming represents a stereotypic form of exploratory movement, which enables these cells to gather information about the direction of concentration gradient. This information is encoded in the relative phase of temporal oscillations of the concentration stimulus. Unlike bacteria that employ a fundamentally stochastic chemotaxis strategy of run-&-tumble, sperm use directed steering responses[8,140]. Generally, helical chemotaxis is expected to be more efficient than a biased random walk. This is because helical chemotaxis enables the cell to align its direction of net motion parallel to the direction of the gradient. Additionally, measuring concentration gradients while moving along helical paths provides an effective mean to integrate out molecular shot noise of chemosensation[141,142].

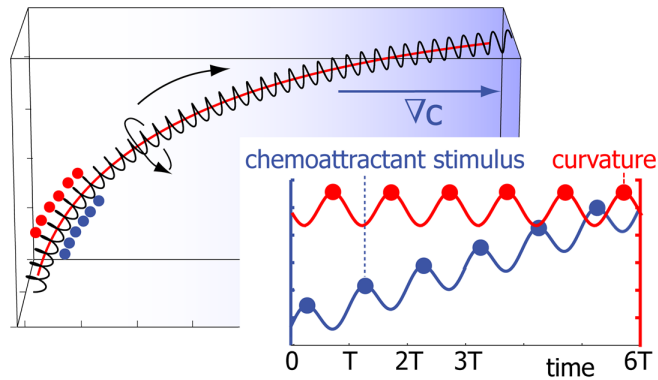

**Figure 9:** *The principle of helical chemotaxis.* We describe a navigation strategy of helical chemotaxis, which is employed *e.g.* by sperm from marine species with external fertilization that swim along helical paths[8,93,140]. In the presence of chemoattractant gradient, helical swimming paths bend in the direction of the gradient, which aligns the helix axis with the gradient direction. This chemotaxis strategy relies on a simple geometric principle: While swimming along a helix, the cell periodically moves up and down the gradient. The cell thus perceives an oscillatory chemoattractant stimulus that oscillates with the period $T$ of helical swimming. The cell responds to this oscillatory chemical signal with oscillations of its path curvature. As a result, the helix bends to align its axis with the gradient.

**Adaptation to spatio-temporal stability of concentration gradients.** Cells must navigate in fluctuating environments. For example, chemical gradients established by diffusion in aqueous environments will be not be perfectly linear, but become distorted by turbulent flows[299,300]. As a second example, concentrations gradients of nutrients such as organic debris in the ocean continuously change due to the dynamics of its production and uptake by other organisms[301,302]. In a theoretical description of bacterial chemotaxis, optimal strategies of chemotactic signaling such as time-scales of signal integration or sensory adaptation were found to depend on sensing noise[303], and thus on the characteristics of concentration gradients. Celani and Vergassola formalized a risk-averse *maximin-* strategy for bacterial chemotaxis, which is adapted to random, short-lived concentration gradients[304]. If concentrations change on a typical time-scale, memory can improve the performance of cellular gradient-sensing. Concepts from control engineering (such as Kalman filters) represent strategies for the robust estimation of concentration gradients[305]. Finally, the availability of extensive memory and computation resources facilitates advanced strategies of gradient search. One example of such an advanced strategy is infotaxis.

For the theoretically proposed mechanism of infotaxis, a hypothetical 'search agent' exploits its full sensation history to compute a detailed likelihood map of the location of a single point source[306]. Put differently, the search agent relies on a cognitive model of environmental variability. The search agent then choses its next movement step such as to maximize the expected reduction in Shannon entropy of this likelihood map. This theoretical mechanism represents a viable solution to the general trade-off choice between *exploration* (*i.e.* active movement geared at gathering additional information about the environment) and *exploitation* (*i.e.* directed movement towards the current estimate of the most-likely target position). Infotaxis was shown to perform even for dilute concentration gradients that are distorted by strongly turbulent flows. For single cells with minimal information processing capabilities, however, navigation strategies that require extensive memory and information processing capabilities such as infotaxis may not be available. Cells have to make optimal use of available resources for sensing, information processing, and memory for spatial navigation[307].

**Concluding remark.** In this introduction, we highlighted the nonlinear physics of cell motility and self-organized pattern formation in biological systems. In particular, we emphasized how non-equilibrium fluctuations and external perturbations affect cellular function. In the selected publications of chapter 2 and 3, these two central themes, nonlinear dynamics and fluctuations, are studied for specific biological systems. The systems under study range from flagellar swimming, steering, and synchronization to cytoskeletal pattern formation and self-scaling morphogen gradients. We combine analytically tractable theoretical descriptions and computational approaches. Thereby, we provide insight into physical mechanisms of biological function. Further, we enable a quantitative comparison of theory and experiment. Ultimately, we seek to use theoretical physics to contribute to the understanding of fundamental principles that render biological dynamics robust in the presence of strong fluctuations and perturbations.

# Bibliography

1. Gray, J. *Ciliary Movements*. (Cambridge Univ. Press, 1928).
2. Brokaw, C. J. Flagellar movement: a sliding filament model. *Science* **178,** 455–462 (1972).
3. Alberts, B. *et al. Molecular Biology of the Cell*. (Garland Science, 2002).
4. Taylor, G. I. Analysis of the swimming of microscopic organisms. *Proc. Roy. Soc. Lond. A* **209,** 447–461 (1951).
5. Geyer, V. F., Jülicher, F., Howard, J. & Friedrich, B. M. Cell-body rocking is a dominant mechanism for flagellar synchronization in a swimming alga. *Proc. Natl. Acad. Sci. U. S. A.* **110,** 18058–18063 (2013).
6. Brumley, D. R., Wan, K. Y., Polin, M. & Goldstein, R. E. Flagellar synchronization through direct hydrodynamic interactions. *eLife* **3,** (2014).
7. Witman, G. B. *Chlamydomonas* phototaxis. *Trends Cell Biol* **3,** 403–408 (1993).
8. Alvarez, L., Friedrich, B. M., Gompper, G. & Kaupp, U. B. The computational sperm cell. *Trends Cell Biol.* **24,** 198–207 (2014).
9. Crenshaw, H. C. A new look at locomotion in microorganisms: Rotating and translating. *Americ. Zool.* **36,** 608–618 (1996).
10. Friedrich, B. M., Fischer-Friedrich, E., Gov, N. & Safran, S. Sarcomeric pattern formation by actin cluster coalescence. *PLoS Comp Biol* **8,** e1002544 (2012).
11. Werner, S. *et al.* Scaling and regeneration of self-organized patterns. *Phys. Rev. Lett.* **114,** 5 (2015).
12. Bray, D. *Cell Movements*. (Garland, 2002).
13. Ben-Isaac, E. *et al.* Effective temperature of red-blood-cell membrane fluctuations. *Phys. Rev. Lett.* **106,** 238103 (2011).
14. Ma, R., Klindt, G. S., Riedel-Kruse, I. H., Jülicher, F. & Friedrich, B. M. Active phase and amplitude fluctuations of flagellar beating. *Phys. Rev. Lett.* **113,** 048101 (2014).
15. Laughlin, S. B. The role of sensory adaptation in the retina. *J. Exp. Biol.* **146,** 39–62 (1989).
16. Smirnakis, S. M., Berry, M. J., Warland, D. K., Bialek, W. & Meister, M. Adaptation of retinal processing to image contrast and spatial scale. *Nature* **386,** 69–73 (1997).
17. Barkai, N. & Leibler, S. Robustness in simple biochemical networks. *Nature* **387,** 913–917 (1997).
18. Kondo, S. & Miura, T. Reaction-diffusion model as a framework for understanding biological pattern formation. *Science* **329,** 1616–1620 (2010).
19. Karsenti, E. Self-organization in cell biology: A Brief History. *Nat. Rev. Mol. Cell Biol.* **9,** 255–262 (2008).
20. Nicolis, G. & Prigogine, I. *Self-Organization in Nonequilibrium Systems*. (Wiley & Sons, 1977).
21. Chrzanowska-Wodnicka, M. & Burridge, K. Rho-stimulated contractility drives the formation of stress fibers and focal adhesions. *J. Cell Biol.* **133,** 1403–1415 (1996).
22. April, E. W. Liquid-crystalline characteristics of the thick filament lattice of striated muscle. *Nature* **257,** 139–141 (1975).
23. Anderson, P. More is different: Broken symmetry and the nature of hierarchial structure of science. *Science* **177,** 393–396 (1972).
24. Nelson, P. *Biological Physics*. (Freeman, 2004).
25. Hodgkin, A. L. & Huxley, A. F. A quantitative description of membrane current and its application to conduction and excitation in nerve. *J. Physiol.* **117,** 500–544 (1952).
26. Süel, G. M., Garcia-Ojalvo, J., Liberman, L. M. & Elowitz, M. B. An excitable gene regulatory circuit induces transient cellular differentiation. *Nature* **440,** 545–50 (2006).
27. Kobayashi, T. J. Implementation of dynamic Bayesian decision making by intracellular kinetics. *Phys. Rev. Lett.* **104,** 228104 (2010).
28. Kruse, K. & Jülicher, F. Oscillations in cell biology. *Curr. Opin. Cell Biol.* **17,** 20–26 (2005).
29. Perkins, T. J. & Swain, P. S. Strategies for cellular decision-making. *Mol. Sys. Biol.* **5,** 326 (2009).
30. Balazsi, G., Van Oudenaarden, A. & Collins, J. J. Cellular decision making and biological noise: From microbes to mammals. *Cell* **144,** 910–925 (2011).
31. Ozbudak, E. M., Thattai, M., Lim, H. N., Shraiman, B. I. & Van Oudenaarden, A. Multistability in the lactose utilization network of *Escherichia coli*. *Nature* **427,** 737–740 (2004).
32. Howard, J. Mechanical signaling in networks of motor and cytoskeletal proteins. *Ann. Rev. Biophys.* **38,** 217–234 (2009).
33. Winfree, A. T. *The geometry of biological time: Interdisciplinary applied mathematics*. *Springer* **24,** (2001).
34. Martin, P., Bozovic, D., Choe, Y. & Hudspeth, A. J. Spontaneous oscillation by hair bundles of the bullfrog's sacculus. *J. Neurosci.* **23,** 4533–48 (2003).

35. Nadrowski, B., Martin, P. & Jülicher, F. Active hair-bundle motility harnesses noise to operate near an optimum of mechanosensitivity. *Proc. Natl. Acad. Sci. U.S.A.* **101,** 12195–200 (2004).
36. Crenshaw, H. C. Negative phototaxis in *Chlamydomonas* is via helical klinotaxis - A new paradigm for phototaxis. *Mol. Biol. Cell.* **7,** 48a (1996).
37. Yoshimura, K. & Kamiya, R. The sensitivity of *Chlamydomonas* photoreceptor is optimized for the frequency of cell body rotation. *Plant Cell Physiol.* **42,** 665–72 (2001).
38. Brokaw, C. J. Thinking about flagellar oscillation. *Cell Motil. Cytoskel.* **66,** 425–436 (2008).
39. Riedel-Kruse, I. H., Hilfinger, A., Howard, J. & Jülicher, F. How molecular motors shape the flagellar beat. *HFSP J.* **1,** 192–208 (2007).
40. Turing, A. M. The chemical basis of morphogenesis. *Phil. Trans. R. Soc. Lond. B* **237,** 37–72 (1952).
41. Gierer, A. & Meinhardt, H. A theory of biological pattern formation. *Kybernetik* **12,** 30–39 (1972).
42. Yoshinaga, N., Joanny, J.-F., Prost, J. & Marcq, P. Polarity patterns of stress fibers. *Phys. Rev. Lett.* **105,** 238103 (2010).
43. Bois, J. S., Jülicher, F. & Grill, S. W. Pattern formation in active fluids. *Phys. Rev. Lett.* **106,** 028103(4) (2011).
44. Engler, A. J. *et al.* Myotubes differentiate optimally on substrates with tissue-like stiffness: pathological implications for soft or stiff microenvironments. *J. Cell Biol.* **166,** 877–87 (2004).
45. Yeung, T. *et al.* Effects of substrate stiffness on cell morphology, cytoskeletal structure, and adhesion. *Cell Motil. Cytoskeleton* **60,** 24–34 (2005).
46. Friedrich, B. M. & Safran, S. A. Nematic order by elastic interactions and cellular rigidity sensing. *Europhys. Lett.* **93,** 28007 (2011).
47. Friedrich, B. M., Buxboim, A., Discher, D. E. & Safran, S. A. Striated acto-myosin fibers can reorganize and register in response to elastic interactions with the matrix. *Biophys. J.* **100,** 2706–2715 (2011).
48. Friedrich, B. M. & Safran, S. How cells feel their substrate: spontaneous symmetry breaking of active surface stresses. *Softmatter* **8,** 3223–3230 (2012).
49. McDonagh, M. J. & Davies, C. T. Adaptive response of mammalian skeletal muscle to exercise with high loads. *Eur. J. Appl. Physiol. Occup. Physiol.* **52,** 139–155 (1984).
50. Turner, C. H. Three rules for bone adapation to mechanical stimuli. *Bone* **23,** 399–407 (1998).
51. Pfeifer, R., Lungarella, M. & Iida, F. Self-organization, embodiment, and biologically inspired robotics. *Science* **318,** 1088–1093 (2007).
52. Alon, U., Surette, M. G., Barkai, N. & Leibler, S. Robustness in chemical networks. *Nature* **397,** 168–171 (1999).
53. Kitano, H. Towards a theory of biological robustness. *Mol. Syst. Biol.* **3,** 137 (2007).
54. Germain, R. N. The art of the probable: system control in the adaptive immune system. *Science* **293,** 240–245 (2001).
55. Murase, M. *The Dynamics of Cellular Motility*. John Wiley & Sons **50,** (1992).
56. Berg, H. & Anderson, R. Bacteria swim by rotating their flagellar filaments. *Nature* **245,** 380 – 382 (1973).
57. Shaevitz, J. W., Lee, J. Y. & Fletcher, D. A. Spiroplasma swim by a processive change in body helicity. *Cell* **122,** 941–945 (2005).
58. Gray, J. The movement of sea-urchin spermatozoa. *J. exp. Biol.* **32,** 775–801 (1955).
59. Pollard, T. D. & Borisy, G. G. Cellular motility driven by assembly and disassembly of actin filaments. *Cell* **112,** 453–465 (2003).
60. Rafelski, S. M. & Theriot, J. A. Crawling toward a unified model cell motility : Spatial and temporal regulation of actin dynamics. *Annu. Rev. Biochem.* **73,** 209–39 (2004).
61. Devreotes, P. N. & Zigmond, S. H. Chemotaxis in eukaryotic cells: a focus on leukocytes and *Dictyostelium*. *Annu. Rev. Cell Biol.* **4,** 649–686 (1988).
62. Rappel, W.-J., Thomas, P. J., Levine, H. & Loomis, W. F. Establishing direction during chemotaxis in eukaryotic cells. *Biophys. J.* **83,** 1361–7 (2002).
63. Howard, J. *Mechanics of Motor Proteins and the Cytoskeleton*. (Sinauer, 2001).
64. Fritzsche, M., Lewalle, A., Duke, T., Kruse, K. & Charras, G. Analysis of turnover dynamics of the submembranous actin cortex. *Mol. Biol. Cell* **24,** 757–67 (2013).
65. Footer, M. J., Kerssemakers, J. W. J., Theriot, J. A. & Dogterom, M. Direct measurement of force generation by actin filament polymerization using an optical trap. *Proc. Natl. Acad. Sci. U.S.A.* **104,** 2181–2186 (2007).
66. Nicastro, D. *et al.* The molecular architecture of axonemes revealed by cryoelectron tomography. *Science* **313,** 944–8 (2006).
67. Shih, Y.-L. & Rothfield, L. The bacterial cytoskeleton. *Microbiol. Mol. Biol. Rev.* **70,** 729–754 (2006).

68. Bugyi, B. & Carlier, M.-F. Control of actin filament treadmilling in cell motility. *Annu. Rev. Biophys.* **39,** 449–470 (2010).
69. Berg, H. C. *E. coli in Motion*. (Springer, 2004).
70. Hunt, A. J., Gittes, F. & Howard, J. The force exerted by a single kinesin molecule against a viscous load. *Biophys. J.* **67,** 766–781 (1994).
71. Yasuda, R., Noji, H., Ishiwata, S., Yoshida, M. & Kinosita Jr., K. $F_1$-ATPase is a highly efficient molecular motor that rotates with discrete 120 steps. *Cell* **93,** 1117–1124 (1998).
72. Itoh, H., Takahashi, A., Adachi, K. & Noji, H. Mechanically driven ATP synthesis by $F_1$-ATPase. *Nature* **427,** 465–468 (2004).
73. Sowa, Y. & Berry, R. M. Bacterial flagellar motor. *Quart. Rev. Biophys.* **41,** 103–132 (2008).
74. Chen, X. & Berg, H. C. Torque-speed relationship of the flagellar rotary motor of *Escherichia coli*. *Biophys. J.* **78,** 1036–1041 (2000).
75. Calladine, C. R. Design requirements for the construction of bacterial flagella. *J. Theor. Biol.* **57,** 469–489 (1976).
76. Srigiriraju, S. V. & Powers, T. R. Continuum model for polymorphism of bacterial flagella. *Phys. Rev. Lett.* **94,** 248101 (2005).
77. Friedrich, B. M. A mesoscopic model for helical bacterial flagella. *J. Math. Biol.* **53,** 162–178 (2006).
78. Vogel, R. & Stark, H. Motor-driven bacterial flagella and buckling instabilities. *Eur. Phys. J. E* **35,** 12015 (2012).
79. Reichert, M. & Stark, H. Synchronization of rotating helices by hydrodynamic interactions. *Eur. Phys. J. E* **17,** 493–500 (2005).
80. Reigh, S. Y., Winkler, R. G. & Gompper, G. Synchronization and bundling of anchored bacterial flagella. *Soft Matter* **8,** 4363 (2012).
81. Berg, H. C. & Brown, D. A. Chemotaxis in *Escherichia coli* analysed by three-dimensional tracking. *Nature* **239,** 500–504 (1972).
82. Vogel, G. Betting on cilia. *Nature* **310,** 216–218 (2005).
83. Marshall, W. F. & Nonaka, S. Cilia: tuning in to the cell's antenna. *Curr. Biol.* **16,** R604–14 (2006).
84. Gibbons, I. R. & Rowe, A. J. Dynein: A protein with adenosine triphosphatase activity from cilia. *Science* **183,** 139–141 (1965).
85. Machin, K. E. Wave propagation along flagella. *J. exp. Biol.* **35,** 796–806 (1958).
86. Afzelius, B. A., R. Dallai, S. L. & Bellon, P. L. Flagellar structure in normal human spermatozoa and in spermatozoa that lack dynein arms. *Tissue Cell* **27,** 241–247 (1995).
87. Carvalho-Santos, Z., Azimzadeh, J., Pereira-Leal, J. B. & Bettencourt-Dias, M. Tracing the origins of centrioles, cilia, and flagella. *J. Cell Biol.* **194,** 165–175 (2011).
88. Summers, K. E. & Gibbons, I. R. Adenosine Triphosphate-Induced Sliding of Tubules in Trypsin-Treated Flagella of Sea-Urchin Sperm. *Proc. Natl. Acad. Sci.* **68,** 3092–3096 (1971).
89. Brokaw, C. J. Direct measurements of sliding between outer doublet microtubules in swimming sperm flagella. *Science* **243,** 1593–6 (1989).
90. Bui, K. H., Sakakibara, H., Movassagh, T., Oiwa, K. & Ishikawa, T. Asymmetry of inner dynein arms and inter-doublet links in *Chlamydomonas* flagella. *J. Cell Biol.* **186,** 437–46 (2009).
91. Crenshaw, H. C., Ciampaglio, C. N. & McHenry, M. Analysis of the three-dimensional trajectories of organisms: Estimates of velocity, curvature and torsion from positional information. *J. exp. Biol.* **203,** 961–982 (2000).
92. Corkidi, G., Taboada, B., Wood, C. D., Guerrero, A. & Darszon, A. Tracking sperm in three-dimensions. *Biochem. Biophys. Res. Comm.* **373,** 125–129 (2008).
93. Jikeli, J. F. *et al.* Sperm navigation along helical paths in 3D chemoattractant landscapes. *Nat. Comm.* **6,** 7985 (2015).
94. Rüffer, U. & Nultsch, W. Flagellar photoresponses of *Chlamydomonas* cells held on micropipettes: II. Change in flagellar beat frequency. *Cell Mot. Cytoskel.* **15,** 162–167 (1991).
95. Jülicher, F. & Prost, J. Spontaneous oscillations of collective molecular motors. *Phys. Rev. Lett.* **78,** 4510–4513 (1997).
96. Brokaw, C. J. Effects of viscosity and ATP concentration on the movement of reactivated sea-urchin sperm flagella. *J. exp. Biol.* **62,** 701–19 (1975).
97. Alper, J., Geyer, V., Mukundan, V. & Howard, J. Reconstitution of flagellar sliding. *Methods Enzymol.* **524,** 343–369 (2013).
98. Geyer, V. F., Sartori, P., Friedrich, B. M., Jülicher, F. & Howard, J. Independent control of the static and dynamic components of the Chlamydomonas flagellar beat. *Curr. Biol.* **26,** 1098–1103 (2016).
99. Brokaw, C. J. Bend propagation by a sliding filament model for flagella. *J. exp. Biol.* **55,** 289–304 (1971).

100. Camalet, S. & Jülicher, F. Self-organized beating and swimming of internally driven filaments. *Phys. Rev. Lett.* **82,** 1590–1593 (1999).
101. Camalet, S. & Jülicher, F. Generic aspects of axonemal beating. *New J. Phys.* **2,** 1–23 (2000).
102. Lindemann, C. B. A 'geometric clutch' hypothesis to explain oscillations of the axoneme of cilia and flagella. *J. Theor. Biol.* **168,** 175–189 (1994).
103. Bayly, P. V. & Wilson, K. S. Analysis of unstable modes distinguishes mathematical models of flagellar motion. *J. R. Soc. Interface* **12,** 20150124–20150124 (2015).
104. Sartori, P., Geyer, V., Scholich, A., Jülicher, F. & Howard, J. Dynamic curvature regulation accounts for the symmetric and asymmetric beats of *Chlamydomonas* flagella. 1–25 (2015).
105. Rüffer, U. & Nultsch, W. High-speed cinematographic analysis of the movement of *Chlamydomonas*. *Cell Mot.* **5,** 251–263 (1985).
106. Broadhead, R. *et al.* Flagellar motility is required for the viability of the bloodstream trypanosome. *Nature* **440,** 224–227 (2006).
107. Krüger, T. & Engstler, M. Flagellar motility in eukaryotic human parasites. *Semin. Cell Dev. Biol.* **46,** 113–127 (2015).
108. Purcell, E. M. Life at low Reynolds numbers. *Am. J. Phys.* **45,** 3–11 (1977).
109. Lauga, E. & Powers, T. R. The hydrodynamics of swimming microorganisms. *Rep. Prog. Phys.* **72,** 096601 (2009).
110. Elgeti, J., Winkler, R. G. & Gompper, G. Physics of microswimmers - single particle motion and collective behavior: a review. *Rep. Prog. Phys.* **78,** 056601 (2015).
111. Sleigh, M. A. *Cilia and Flagella*. (Academic Press, 1974).
112. Vilfan, A. & Jülicher, F. Hydrodynamic flow patterns and synchronization of beating cilia. *Phys. Rev. Lett.* **96,** 58102 (2006).
113. Golestanian, R. & Ajdari, A. Analytic results for the three-sphere swimmer at low Reynolds number. *Phys. Rev. E* **77,** 036308 (2008).
114. Dreyfus, R., Baudry, J. & Stone, H. A. Purcell's rotator: mechanical rotation at low Reynolds number. *Eur. Phys. J. B* **47,** 161–164 (2005).
115. Polotzek, K. & Friedrich, B. M. A three-sphere swimmer for flagellar synchronization. *New J. Phys.* **15,** 045005 (2013).
116. Klindt, G. S. & Friedrich, B. M. Flagellar swimmers oscillate between pusher- and puller-type swimming. *Phys. Rev. E* **92,** 063019 (2015).
117. Liu, Y. J. *Fast Multipole Boundary Element Method: Theory and Applications in Engineering*. (Cambridge Univ. Press, 2009).
118. Brokaw, C. J. Effects of increased viscosity on the movements of some invertebrate spermatozoa. *J. exp. Biol.* **45,** 113–139 (1966).
119. Friedrich, B. M., Riedel-Kruse, I. H., Howard, J. & Jülicher, F. High-precision tracking of sperm swimming fine structure provides strong test of resistive force theory. *J. exp. Biol.* **213,** 1226–1234 (2010).
120. Shapere, A. & Wilczek, F. Self-propulsion at low Reynolds number. *Phys. Rev. Lett.* **58,** 2051–2054 (1987).
121. Werner, S., Rink, J. C., Riedel-Kruse, I. H. & Friedrich, B. M. Shape mode analysis exposes movement patterns in biology: Flagella and flatworms as case studies. *PLoS One* **9,** e113083 (2014).
122. Elfring, G. & Lauga, E. Hydrodynamic phase locking of swimming microorganisms. *Phys. Rev. Lett.* **103,** 088101 (2009).
123. Singla, V. & Reiter, J. F. The primary cilium as the cell's antenna: signaling at a sensory organelle. *Science* **313,** 629–33 (2006).
124. Brennen, C. & Winet, H. Fluid mechanics of propulsion by cilia and flagella. *Annu. Rev. Fluid Mech.* **9,** 339–398 (1977).
125. Sanderson, M. J. & Sleigh, M. A. Ciliary activity of cultured rabbit tracheal epithelium: beat pattern and metachrony. *J. Cell Sci.* **47,** 331–47 (1981).
126. Worthington, W. C. & Cathcart, R. S. Ependymal cilia: Distribution and activity in the adult human brain. *Science* **139,** 221–222 (1963).
127. Cartwright, J. H. E., Piro, O. & Tuval, I. Fluid-dynamical basis of the embryonic development of left-right asymmetry in vertebrates. *Proc. Natl. Acad. Sci. U.S.A.* **101,** 7234–9 (2004).
128. Freund, J. B., Goetz, J. G., Hill, K. L. & Vermot, J. Fluid flows and forces in development: functions, features and biophysical principles. *Development* **139,** 3063–3063 (2012).
129. Boehmer, M. *et al.* $Ca^{2+}$ spikes in the flagellum control chemotactic behavior of sperm. *Eur. Mol. Biol. Organ. J.* **24,** 2741–2752 (2005).
130. Wood, C. D., Nishigaki, T., Furuta, T., Baba, S. A. & Darszon, A. Real-time analysis of the role of $Ca^{2+}$ in flagellar movement and motility in single sea urchin sperm. *J. Cell Biol.* **169,** 725–31 (2005).

131. Hilfinger, A. & Jülicher, F. The chirality of ciliary beats. *Phys. Biol.* **5,** 016003 (2008).
132. Wilson, L. G., Carter, L. M. & Reece, S. E. High-speed holographic microscopy of malaria parasites reveals ambidextrous flagellar waveforms. *Proc Natl Acad Sci U S A* **110,** 18769–18774 (2013).
133. Ishijima, S., Oshio, S. & Mohri, H. Flagellar movement of human spermatozoa. *Gamete Res.* **13,** 185–197 (1986).
134. Su, T. *et al.* Sperm trajectories form chiral ribbons. *Sci. Rep.* **3,** 1664 (2013).
135. Rothschild, L. Non-random distribution of bull spermatozoa in a drop of sperm suspension. *Nature* **198,** 1221–1222 (1963).
136. Elgeti, J., Kaupp, U. B. & Gompper, G. Hydrodynamics of sperm cells near surfaces. *Biophys. J.* **99,** 1018–26 (2010).
137. Elgeti, J. & Gompper, G. Self-propelled rods near surfaces. *Europhys. Lett.* **85,** 38002 (2009).
138. Smith, D. J., Gaffney, E. A., Blake, J. R. & Kirkman-Brown, J. C. Human sperm accumulation near surfaces: a simulation study. *J. Fluid Mech.* **621,** 289–320 (2009).
139. Crenshaw, H. C. Orientation by helical motion - III. Microorganisms can orient to stiumuli by changing the direction of their rotational velocity. *Bull. Math. Biol.* **55,** 231–255 (1993).
140. Friedrich, B. M. & Jülicher, F. Chemotaxis of sperm cells. *Proc. Natl. Acad. Sci. U. S. A.* **104,** 13256–13261 (2007).
141. Friedrich, B. M. & Jülicher, F. The stochastic dance of circling sperm cells: Sperm chemotaxis in the plane. *New J. Phys.* **10,** 123025(19) (2008).
142. Friedrich, B. M. & Jülicher, F. Steering chiral swimmers along noisy helical paths. *Phys. Rev. Lett.* **103,** 68102 (2009).
143. Eisenbach, M. & Giojalas, L. C. Sperm guidance in mammals - an unpaved road to the egg. *Nat. Rev. Mol. Cell Biol.* **7,** 276–285 (2006).
144. Kaupp, U. B., Hildebrand, E. & Weyand, I. Sperm Chemotaxis in Marine Invertebrates — Molecules and Mechanisms. *J. Cell. Physiol.* **208,** 487–494 (2006).
145. Zacks, D. N. & Spudich, J. L. Gain setting in *Chlamydomonas reinhardtii*: mechanism of phototaxis and the role of the photophobic response. *Cell Motil. Cytoskeleton* **29,** 225–30 (1994).
146. Nonaka, S., Shiratori, H., Saijoh, Y. & Hamada, H. Determination of left-right patterning of the mouse embryo by artificial nodal flow. *Nature* **418,** 96–99 (2002).
147. Winet, H., Bernstein, G. S. & Head, J. Observations on the response of human spermatozoa to gravity, boundaries and fluid shear. *J. Reprod. Fert.* **70,** 511–523 (1984).
148. Miki, K. & Clapham, D. E. Rheotaxis guides mammalian sperm. *Curr. Biol.* **23,** 443–52 (2013).
149. Kantsler, V., Dunkel, J., Blayney, M. & Goldstein, R. E. Rheotaxis facilitates upstream navigation of mammalian sperm cells. *eLife* **3,** e02403 (2014).
150. Heddergott, N. *et al.* Trypanosome motion represents an adaptation to the crowded environment of the vertebrate bloodstream. *PLoS Pathog.* **8,** e1003023 (2012).
151. Wan, K. Y. & Goldstein, R. E. Rhythmicity, recurrence and recovery of flagellar beating. *Phys. Rev. Lett.* **113,** 238103 (2014).
152. Qin, B., Gopinath, A., Yang, J., Gollub, J. P. & Arratia, P. E. Flagellar kinematics and swimming of algal cells in viscoelastic fluids. *Sci. Rep.* **5,** 9190 (2015).
153. Chen, D. T. N., Heymann, M., Fraden, S., Nicastro, D. & Dogic, Z. ATP consumption of eukaryotic flagella measured at a single-cell level. *Biophys. J.* **109,** 2562–2573 (2015).
154. Woolley, D. M., Crockett, R. F., Groom, W. D. I. & Revell, S. G. A study of synchronisation between the flagella of bull spermatozoa, with related observations. *J. exp. Biol.* **212,** 2215–23 (2009).
155. Okuno, M. & Hiramoto, Y. Mechanical stimulation of starfish sperm flagella. *J. exp. Biol.* **65,** 401–13 (1976).
156. Quaranta, G., Aubin-Tam, M. E. & Tam, D. Hydrodynamics versus intracellular coupling in the synchronization of eukaryotic flagella. *Phys. Rev. Lett.* **115,** 238101 (2015).
157. Osterman, N. & Vilfan, A. Finding the ciliary beating pattern with optimal efficiency. *Proc. Natl. Acad. Sci. U.S.A.* **108,** 15727–32 (2011).
158. Elgeti, J. & Gompper, G. Emergence of metachronal waves in cilia arrays. *Proc. Natl. Acad. Sci. U.S.A.* **110,** 4470–5 (2013).
159. Rüffer, U. & Nultsch, W. Flagellar coordination in *Chlamydomonas* cells held on micropipettes. *Cell Motil. Cytoskel.* **41,** 297–307 (1998).
160. Goldstein, R. E., Polin, M. & Tuval, I. Noise and synchronization in pairs of beating eukaryotic flagella. *Phys. Rev. Lett.* **103,** 168103 (2009).
161. Goldstein, R. E., Polin, M. & Tuval, I. Emergence of synchronized beating during the regrowth of eukaryotic flagella. *Phys. Rev. Lett.* **107,** 148103 (2011).
162. Tam, D. & Hosoi, A. E. Optimal feeding and swimming gaits of biflagellated organisms. *Proc. Natl. Acad. Sci. U.S.A.* **108,** 1001–6 (2011).

163. Riedel, I. H., Kruse, K. & Howard, J. A self-organized vortex array of hydrodynamically entrained sperm cells. *Science* **309,** 300–303 (2005).
164. Shiba, K. *et al.* Sperm-activating peptide induces asymmetric flagellar bending in sea urchin sperm. *Zoolog. Sci.* **22,** 293–9 (2005).
165. Josef, K., Saranak, J. & Foster, K. W. Ciliary behaviour of a negatively phototactic *Chlamydomonas reinhardtii*. *Cell Mot. Cytoskel.* **61,** 97–111 (2005).
166. Fujiu, K., Nakayama, Y., Iida, H., Sokabe, M. & Yoshimura, K. Mechanoreception in motile flagella of *Chlamydomonas*. *Nat. Cell Biol.* **13,** 630–632 (2011).
167. Suarez, S. S. Control of hyperactivation in sperm. *Hum. Reprod. Update* **14,** 647–657 (2008).
168. Brokaw, C. J. Calcium-induced asymmetrical beating of triton-demembranated sea urchin sperm flagella. *J. Cell Biol.* **82,** 401–411 (1979).
169. Alvarez, L. *et al.* The rate of change in $Ca^{2+}$ concentration controls sperm chemotaxis. *J Cell Biol* **196,** (2012).
170. Sakato, M. & King, S. M. Calcium regulates ATP-sensitive microtubule binding by *Chlamydomonas* outer arm dynein. *J. Biol. Chem.* **278,** 43571–43579 (2003).
171. DiPetrillo, C. G. & Smith, E. F. Pcdp1 is a central apparatus protein that binds $Ca^{2+}$-calmodulin and regulates ciliary motility. *J. Cell Biol.* **189,** 601–612 (2010).
172. Mizuno, K. *et al.* Calaxin drives sperm chemotaxis by $Ca^{2+}$-mediated direct modulation of a dynein motor. *Proc. Natl. Acad. Sci. U. S. A.* **109,** 20497–502 (2012).
173. Turner, R. M. Moving to the beat: A review of mammalian sperm motility regulation. *Reprod. Fertil. Dev.* **18,** 25–38 (2006).
174. Merz, A. J., So, M. & Sheetz, M. P. Pilus retraction powers bacterial twitching motility. *Nature* **407,** 98–102 (2000).
175. Skerker, J. M. & Berg, H. C. Direct observation of extension and retraction of type IV pili. *Proc. Natl. Acad. Sci. U. S. A.* **98,** 6901–6904 (2001).
176. Maier, B. & Wong, G. C. L. How bacteria use type IV pili machinery on surfaces. *Trends Microbiol.* **23,** 775–788 (2015).
177. Zaburdaev, V. *et al.* Uncovering the mechanism of trapping and cell orientation during *Neisseria gonorrhoeae* twitching motility. *Biophys. J.* **107,** 1523–1531 (2014).
178. Friedl, P., Borgmann, S. & Bröcker, E. B. Amoeboid leukocyte crawling through extracellular matrix: lessons from the *Dictyostelium* paradigm of cell movement. *J. Leukoc. Biol.* **70,** 491–509 (2001).
179. Thompson, D. W. *On Growth and Form*. (Cambridge Univ. Press, 1942).
180. Berg, H. C. The rotary motor of bacterial flagella. *Ann. Rev. Biochem.* **72,** 19–54 (2003).
181. Macnab, R. M. How bacteria assemble flagella. *Annu. Rev. Microbiol.* **57,** 77–100 (2003).
182. Rosenbaum, J. L. & Witman, G. B. Intraflagellar transport. *Nat. Rev. Mol. Cell Biol.* **3,** 813–825 (2002).
183. Stepanek, L. & Pigino, G. Microtubule doublets are double-track railways for intraflagellar transport trains. *Science* **352,** 721–724 (2016).
184. Nédélec, F. J., Surrey, T., Maggs, a C. & Leibler, S. Self-organization of microtubules and motors. *Nature* **389,** 305–8 (1997).
185. Schaller, V., Weber, C., Semmrich, C., Frey, E. & Bausch, A. R. Polar patterns of driven filaments. *Nature* **467,** 73–77 (2010).
186. Schaller, V., Weber, C. a, Hammerich, B., Frey, E. & Bausch, A. R. Frozen steady states in active systems. *Proc. Natl. Acad. Sci. U. S. A.* **108,** 19183–8 (2011).
187. Sanchez, T., Welch, D., Nicastro, D. & Dogic, Z. Cilia-Like Beating of Active Microtubule Bundles. *Science* **333,** 456 –459 (2011).
188. Dunkel, J. *et al.* Fluid dynamics of bacterial turbulence. *Phys. Rev. Lett.* **110,** 228102 (2013).
189. Sokolov, A. & Aranson, I. S. Physical properties of collective motion in suspensions of bacteria. *Phys. Rev. Lett.* **109,** 248109 (2012).
190. Vicsek, T., Czirok, A., Ben-Jacob, E., Cohen, I. & Shochet, O. Novel type of phase transition in a system of self propelled particles. *Phys. Rev. Lett.* **75,** 1226–1229 (1995).
191. Toner, J. & Tu, Y. H. Long-range order in a 2-dimensional dynamical XY model: How birds fly together. *Phys. Rev. Lett.* **75,** 4326–4329 (1995).
192. Chaté, H., Ginelli, F., Gregoire, G., Peruani, F. & Raynaud, F. Modeling collective motion: Variations on the Vicsek model. *Eur. Phys. J. B* **64,** 451–456 (2008).
193. Marchetti, M. C. *et al.* Hydrodynamics of soft active matter. *Rev. Mod. Phys.* **85,** 1143–1189 (2013).
194. Loose, M. *et al.* Spatial regulators for bacterial cell division self-organize into surface waves in vitro. *Science* **320,** 789–792 (2008).
195. Sheth, R. *et al.* Hox genes regulate digit patterning by controlling the wavelength of a Turing-type mechanism. *Science* **338,** 1476–1480 (2012).

196. Yamaguchi, M., Yoshimoto, E. & Kondo, S. Pattern regulation in the stripe of zebrafish suggests an underlying dynamic and autonomous mechanism. *Proc. Natl. Acad. Sci. U. S. A.* **104,** 4790–4793 (2007).
197. Watanabe, M. & Kondo, S. Is pigment patterning in fish skin determined by the Turing mechanism? *Trends Genet.* **31,** 88–96 (2015).
198. Winfree, A. T. The prehistory of the Belousov-Zhabotinsky oscillator. *J. Chem. Educ.* **61,** 661 (1984).
199. Backouche, F., Haviv, L., Groswasser, D. & Bernheim-Groswasser, A. Active gels: dynamics of patterning and self-organization. *Phys. Biol.* **3,** 264–273 (2006).
200. Soares e Silva, M. *et al.* Active multistage coarsening of actin networks driven by myosin motors. *Proc. Natl. Acad. Sci. U.S.A.* **108,** 9408–13 (2011).
201. Köhler, S., Schaller, V. & Bausch, A. R. Structure formation in active networks. *Nat. Mater.* **10,** 462–468 (2011).
202. Smith, D. *et al.* Molecular motor-induced instabilities and cross linkers determine biopolymer organization. *Biophys. J.* **93,** 4445–4452 (2007).
203. Gorfinkiel, N. & Blanchard, G. B. Dynamics of actomyosin contractile activity during epithelial morphogenesis. *Curr. Opin. Cell Biol.* **23,** 531–539 (2011).
204. Pullarkat, P. a., Fernández, P. a. & Ott, A. Rheological properties of the eukaryotic cell cytoskeleton. *Phys. Rep.* **449,** 29–53 (2007).
205. Kruse, K. & Jülicher, F. Actively contracting bundles of polar filaments. *Phys. Rev. Lett.* **85,** 1778–81 (2000).
206. Lenz, M., Thoresen, T., Gardel, M. L. & Dinner, A. R. Contractile units in disordered actomyosin bundles arise from F-actin buckling. *Phys. Rev. Lett.* **108,** 238107 (2012).
207. Kruse, K., Zumdieck, A. & Jülicher, F. Continuum theory of contractile fibres. *Eur. Phys. Lett.* **64,** 716–722 (2003).
208. Martin, P. & Lewis, J. Actin cables and epidermal movement in embryonic wound healing. *Nature* **360,** 179–183 (1992).
209. Peterson, L. J. *et al.* Simultaneous stretching and contraction of stress fibers in vivo. *Mol. Biol. Cell* **15,** 3497–3508 (2004).
210. Roland, J., Berro, J., Michelot, A., Blanchoin, L. & Martiel, J.-L. Stochastic severing of actin filaments by actin depolymerizing factor/cofilin controls the emergence of a steady dynamical regime. *Biophys. J.* **94,** 2082–94 (2008).
211. Gregorio, C. C., Granzier, H., Sorimachi, H. & Labeit, S. Muscle assembly: A titanic achievement? *Curr. Opin. Cell Biol.* **11,** 18–25 (1999).
212. Ingber, D. Tensegrity: The architectural basis of cellular mechanotransduction. *Ann. Rev. Phys.* **59,** 575–599 (1997).
213. Hyman, A. a & Karsenti, E. Morphogenetic properties of microtubules and mitotic spindle assembly. *Cell* **84,** 401–410 (1996).
214. Solari, C. A., Ganguly, S., Kessler, J. O., Michod, R. E. & Goldstein, R. E. Multicellularity and the functional interdependence of motility and molecular transport. *Proc. Natl. Acad. Sci. U.S.A.* **103,** 1353–1358 (2006).
215. Svitinka, T., Verkhovsky, A. B. & Borisy, G. G. Improved procedures for electron microscopic visualization of the cytoskeleton of cultured cells. *J Struct Biol* **115,** 290–303 (1995).
216. Paulsen, G. *et al.* Subcellular movement and expression of HSP27, B-crystallin, and HSP70 after two bouts of eccentric exercise in humans. *J Appl Physiol* **107,** 570–582 (2009).
217. Iwamoto, H., Niskikawa, Y., Wakayama, J. & Fujisawa, T. Direct X-ray observations of a single hexagonal myofilament lattice in native myofibrils of striated muscle. *Biophys. J.* **83,** 1074–1081 (2002).
218. Rhee, D., Sanger, J. M. & Sanger, J. W. The premyofibril: evidence for its role in myofibrillogenesis. *Cell Motil. Cytoskeleton* **28,** 1–24 (1994).
219. Sanger, J. W. *et al.* How to build a myofibril. *J. Muscle Res. Cell Mot.* **26,** 343–54 (2005).
220. Borisov, A. B., Martynova, M. G. & Russell, M. W. Early incorporation of obscurin into nascent sarcomeres: implication for myofibril assembly during cardiac myogenesis. *Histochem. Cell Biol.* **129,** 463–78 (2008).
221. Weitkunat, M., Kaya-Copur, A., Grill, S. W. & Schnorrer, F. Tension and force-resistant attachment are essential for myofibrillogenesis in drosophila flight muscle. *Curr. Biol.* **24,** 705–716 (2014).
222. Zemel, A. & Mogilner, A. Motor-induced sliding of microtubule and actin bundles. *Phys. Chem. Chem. Phys.* **11,** 4821–4833 (2009).
223. Wolpert, L. & Tickle, C. *Principles of Development*. (Oxford Univ. Press, 2011).
224. Niehrs, C. On growth and form: a Cartesian coordinate system of Wnt and BMP signaling specifies bilaterian body axes. *Development* **137,** 845–857 (2010).

225. Spencer, F. A., Hoffmann, F. M. & Gelbart, W. M. Decapentaplegic: A gene complex affecting morphogenesis in *Drosophila melanogaster*. *Cell* **28,** 451–461 (1982).
226. Wolpert, L. Positional information and the spatial pattern of cellular differentiation. *J. Theor. Biol.* **25,** 1–47 (1969).
227. Martinez Arias, A. Wnts as morphogens? The view from the wing of *Drosophila*. *Nat. Rev. Mol. Cell Biol.* **4,** 321–325 (2003).
228. Kruse, K., Pantazis, P., Bollenbach, T., Jülicher, F. & González-Gaitán, M. Dpp gradient formation by dynamin-dependent endocytosis: Receptor trafficking and the diffusion model. *Development* **131,** 4843–56 (2004).
229. Wartlick, O. *et al.* Dynamics of Dpp signaling and proliferation control. *Science* **331,** 1154–1159 (2011).
230. Ben-Zvi, D., Pyrowolakis, G., Barkai, N. & Shilo, B. Z. Expansion-repression mechanism for scaling the Dpp activation gradient in *Drosophila* wing imaginal discs. *Curr. Biol.* **21,** 1391–1396 (2011).
231. Hamaratoglu, F., de Lachapelle, A. M., Pyrowolakis, G., Bergmann, S. & Affolter, M. Dpp signaling activity requires pentagone to scale with tissue size in the growing *Drosophila* wing imaginal disc. *PLoS Biol* **9,** (2011).
232. Othmer, H. G. H. G. & Pate, E. E. Scale-invariance in reaction-diffusion models of spatial pattern formation. *Proc. Natl. Acad. Sci. U.S.A.* **77,** 4180–4184 (1980).
233. Ben-Zvi, D. & Barkai, N. Scaling of morphogen gradients by an expansion-repression integral feedback control. *Proc. Natl. Acad. Sci. U.S.A.* **107,** 6924–9 (2010).
234. Wartlick, O., Mumcu, P., Jülicher, F. & Gonzalez-Gaitan, M. Understanding morphogenetic growth control - lessons from flies. *Nat. Rev. Mol. Cell Biol.* **12,** 594–604 (2011).
235. Newmark, P. A. & Sánchez Alvarado, A. Not your father's planarian: A classic model enters the era of functional genomics. *Nat. Rev. Genet.* **3,** 210–9 (2002).
236. Gurley, K. A., Rink, J. C. & Alvarado, A. S. beta-catenin defines head versus tail identity during Planarian regeneration and homeostasis. *Science* **319,** 323–327 (2008).
237. Frey, E. & Kroy, K. Brownian motion: a paradigm of soft matter and biological physics. *Ann. Phys.* **14,** 20–50 (2005).
238. Van Kampen, N. G. *Stochastic Processes in Physics and Chemistry*. (North Holland, 1992).
239. Guptasarma, P. Does replication-induced transcription regulate synthesis of the myriad low copy number proteins of *Escherichia coli*? *Bioessays* **17,** 987–997 (1995).
240. Elowitz, M. B., Levine, A. J., Siggia, E. D. & Swain, P. S. Stochastic gene expression in a single cell. **297,** 1183–1187 (2002).
241. Kussell, E. & Leibler, S. Phenotypic diversity, population growth, and information in fluctuating environments. *Science* **309,** 2075–8 (2005).
242. Carter, N. J. & Cross, R. a. Mechanics of the kinesin step. *Nature* **435,** 308–12 (2005).
243. Dubuis, J. O., Tkacik, G., Wieschaus, E. F., Gregor, T. & Bialek, W. Positional information, in bits. *Proc. Natl. Acad. Sci. U.S.A.* **110,** 16301–16308 (2013).
244. Bialek, W. Physical limits to sensation and perception. *Ann Rev. Biophys. Biochem. Chem.* **16,** 455–78 (1987).
245. Fuortes, M. G. & S., Y. Probability of occurrence of discrete potential waves in the eye of Limulus. *J. Gen. Physiol.* **47,** 443–63 (1964).
246. Baylor, B. Y. D. a, Lamb, T. D. & Yau, K. Responses of retinal rods to single photons. *J. Physiol.* **288,** 613–634 (1979).
247. Nelson, P. C. Old and new results about single-photon sensitivity in human vision. *Phys. Biol.* **13,** 025001 (2016).
248. Menini, A., Picco, C. & Firestein, S. Quantal-like current fluctuations induced by odorants in olfactory receptor cells. *Nature* **373,** 435–437 (1995).
249. Strünker, T., Alvarez, L. & Kaupp, U. B. At the physical limit - chemosensation in sperm. *Curr. Opin. Neurobiol.* **34,** 110–116 (2015).
250. Bialek, W. & Setayeshgar, S. Physical limits to biochemical signaling. *Proc. Natl. Acad. Sci. U.S.A.* **102,** 10040–10045 (2005).
251. Govern, C. C. & Ten Wolde, P. R. Energy dissipation and noise correlations in biochemical sensing. *Phys. Rev. Lett.* **113,** 258102 (2014).
252. Finer, J. T., Simmons, R. M. & Spudich, J. A. Single myosin molecule mechanics: piconewton forces and nanometre steps. *Nature* **367,** 566 – 568 (1994).
253. Kubo, R. The fluctuation-dissipation theorem. *Rep. Prog. Phys.* **29,** 255 (1966).
254. Betz, T., Lenz, M., Joanny, J.-F. & Sykes, C. ATP-dependent mechanics of red blood cells. *Proc. Natl. Acad. Sci. U.S.A.* **106,** 15320–5 (2009).

255. Mizuno, D., Tardin, C., Schmidt, C. F. & Mackintosh, F. C. Nonequilibrium mechanics of active cytoskeletal networks. *Science* **315,** 370–373 (2007).
256. Brangwynne, C. P., Koenderink, G. H., MacKintosh, F. C. & Weitz, D. a. Nonequilibrium microtubule fluctuations in a model cytoskeleton. *Phys. Rev. Lett.* **100,** 118104 (2008).
257. Otten, M. *et al.* Local motion analysis reveals impact of the dynamic cytoskeleton on intracellular subdiffusion. *Biophys. J.* **102,** 758–67 (2012).
258. Martin, P., Hudspeth, A. J. & Jülicher, F. Comparison of a hair bundle's spontaneous oscillations with its response to mechanical stimulation reveals the underlying active process. *Proc. Natl. Acad. Sci. U.S.A.* **98,** 14380–14385 (2001).
259. Polin, M., Tuval, I., Drescher, K., Gollub, J. P. & Goldstein, R. E. *Chlamydomonas* swims with two 'gears' in a eukaryotic version of run-and-tumble. *Science* **325,** 487–490 (2009).
260. Drescher, K., Dunkel, J., Cisneros, L. H., Ganguly, S. & Goldstein, R. E. Fluid dynamics and noise in bacterial cell–cell and cell–surface scattering. *Proc. Natl. Acad. Sci. U.S.A.* **108,** 10940(6) (2011).
261. Endow, S. A. & Higuchi, H. A mutant of the motor protein kinesin that moves in both directions on microtubules. *Nature* **406,** 913–6 (2000).
262. Badoual, M., Jülicher, F. & Prost, J. Bidirectional cooperative motion of molecular motors. *Proc. Natl. Acad. Sci. U. S. A.* **99,** 6696–701 (2002).
263. Plaçais, P.-Y., Balland, M., Guérin, T., Joanny, J.-F. & Martin, P. Spontaneous oscillations of a minimal actomyosin system under elastic loading. *Phys. Rev. Lett.* **103,** 158102 (2009).
264. Pikovsky, A. *et al. Synchronization: A Universal Concept in Nonlinear Sciences*. Cambridge Nonlinear Science Series 12 (Cambridge Univ. Press, 2003).
265. Adler, R. Locking phenomena in oscillators. *Proc. IRE* **34,** 351–357 (1946).
266. Stratonovich, R. L. *Topics in the Theory of Random Noise*. (Gordon & Breach, 1963).
267. Pourquié, O. The segmentation clock: converting embryonic time into spatial pattern. *Science* **301,** 328–330 (2003).
268. Strogatz, S. H., Abrams, D. M., McRobie, A., Eckhardt, B. & Ott, E. Theoretical mechanics: crowd synchrony on the Millennium Bridge. *Nature* **438,** 43–44 (2005).
269. Hanggi, P., Talkner, P. & Borkovec, M. Reaction-rate theory - 50 years after Kramers. *Rev. Mod. Phys.* **62,** 251–341 (1990).
270. Ringo, D. L. Flagellar motion and the fine structure of the flagellar apparatus in *Chlamydomonas*. *Cell* **33,** 543–571 (1967).
271. Leptos, K. C. *et al.* Antiphase synchronization in a flagellar-dominance mutant of Chlamydomonas. *Phys. Rev. Lett.* **111,** 158101 (2013).
272. Wan, K. Y., Leptos, K. C. & Goldstein, R. E. Lag, lock, sync, slip : The many 'phases' of coupled flagella. *J Roy Soc Interface* **11,** 20131160 (2014).
273. Friedrich, B. M. Hydrodynamic synchronization of flagellar oscillators. *EPJ Spec. Top. 'Microswimmers'* (editorial recommendation for acceptance) 16 pages (2016).
274. Kotar, J., Leoni, M., Bassetti, B., Lagomarsino, M. C. & Cicuta, P. Hydrodynamic synchronization of colloidal oscillators. *Proc. Natl. Acad. Sci. U.S.A.* **107,** 7669–73 (2010).
275. Bruot, N., Damet, L., Kotar, J., Cicuta, P. & Lagomarsino, M. Noise and synchronization of a single active colloid. *Phys. Rev. Lett.* **107,** 094101 (2011).
276. Bruot, N., Kotar, J., de Lillo, F., Cosentino Lagomarsino, M. & Cicuta, P. Driving potential and noise level determine the synchronization state of hydrodynamically coupled oscillators. *Phys. Rev. Lett.* **109,** 164103 (2012).
277. Di Leonardo, R. *et al.* Hydrodynamic synchronization of light driven microrotors. *Phys. Rev. Lett.* **109,** 034104 (2012).
278. Lhermerout, R., Bruot, N., Cicuta, G. M., Kotar, J. & Cicuta, P. Collective synchronization states in arrays of driven colloidal oscillators. *New J. Phys.* **14,** 105023 (2012).
279. Niedermayer, T., Eckhardt, B. & Lenz, P. Synchronization, phase locking, and metachronal wave formation in ciliary chains. *Chaos* **18,** 037128 (2008).
280. Uchida, N. & Golestanian, R. Generic conditions for hydrodynamic synchronization. *Phys. Rev. Lett.* **106,** 058104 (2011).
281. Golestanian, R., Yeomans, J. M. & Uchida, N. Hydrodynamic synchronization at low Reynolds number. *Soft Matter* **7,** 3074 (2011).
282. Theers, M. & Winkler, R. G. Synchronization of rigid microrotors by time-dependent hydrodynamic interactions. *Phys. Rev. E* **88,** 023012 (2013).
283. Friedrich, B. M. & Jülicher, F. Flagellar synchronization independent of hydrodynamic interactions. *Phys. Rev. Lett.* **109,** 138102 (2012).
284. Berg, H. C. & Purcell, E. M. Physics of chemoreception. *Biophys. J.* **20,** 193–219 (1977).
285. Berg, H. C. *Random Walks in Biology*. (Princeton Univ. Press, 1993).

286. Taylor, G. I. Diffusion by continous movements. *Proc. Roy. Soc. A* **20,** 196–212 (1920).
287. Friedrich, B. M. Search along persistent random walks. *Phys. Biol.* **5,** 026007(6) (2008).
288. Dusenbery, D. B. Minimum size limit for useful locomotion by free-swimming microbes. *Proc. Natl. Acad. Sci. U.S.A.* **94,** 10949–10954 (1997).
289. Tu, Y., Shimizu, T. S. & Berg, H. C. Modeling the chemotactic response of Escherichia coli to time-varying stimuli. *Proc. Natl. Acad. Sci. U. S. A.* **105,** 14855–60 (2008).
290. Kollmann, M., Løvdok, L., Bartholomé, K., Timmer, J. & Sourjik, V. Design principles of a bacterial signalling network. *Nature* **438,** 504–507 (2005).
291. Crenshaw, H. C. Helical orientation - A novel mechanism for the orientation of microorganisms. *Lect. Notes Biophys.* **89,** 361–386 (1990).
292. Levine, H., Kessler, D. A. & Rappel, W. Directional sensing in eukaryotic chemotaxis: A balanced inactivation model. *Proc. Natl. Acad. Sci. U.S.A.* **103,** 9761–9766 (2006).
293. Segall, J. E., Block, S. M. & Berg, H. C. Temporal comparisons in bacterial chemotaxis. *Proc. Natl. Acad. Sci. U.S.A.* **83,** 8987–8991 (1986).
294. Song, L. *et al.* Dictyostelium discoideum chemotaxis: threshold for directed motion. *Eur. J. Cell Biol.* **85,** 981–9 (2006).
295. Amselem, G., Theves, M., Bae, A., Beta, C. & Bodenschatz, E. Control parameter description of eukaryotic chemotaxis. *Phys. Rev. Lett.* **109,** 108103 (2012).
296. Miller, R. L. Sperm chemo-orientation in the metazoa. *Biol. Fertil.* **2,** 275–337 (1985).
297. Kaupp, U. B. *et al.* The signal flow and motor response controling chemotaxis of sea urchin sperm. *Nat. Cell Biol.* **5,** 109–117 (2003).
298. Kashikar, N. D. *et al.* Temporal sampling, resetting, and adaptation orchestrate gradient sensing in sperm. *J. Cell Biol.* **198,** 1075–91 (2012).
299. Riffell, J. A. & Krug, P. J. The ecological and evolutionary consequences of sperm chemoattraction. *Proc. Natl. Acad. Sci. U.S.A.* **101,** 4501–4506 (2004).
300. Stocker, R. Marine microbes see a sea of gradients. *Science* **338,** 628–633 (2012).
301. Kiørboe, T. & Jackson, G. A. Marine snow, organic solute plumes, and optimal chemosensory behavior of bacteria. *Limnol. Oceanogr.* **46,** 1309–1318 (2001).
302. Taylor, J. R. & Stocker, R. Trade-Offs of Chemotactic Foraging in Turbulent Water. *Science* **338,** 675–679 (2012).
303. Andrews, B. W., Yi, T.-M. & Iglesias, P. A. Optimal noise filtering in the chemotactic response of Escherichia coli. *PLoS Comp. Biol.* **2,** e154 (2006).
304. Celani, A. & Vergassola, M. Bacterial strategies for chemotaxis response. *Proc. Natl. Acad. Sci. U.S.A.* **107,** 1391–1396 (2010).
305. Aquino, G., Tweedy, L., Heinrich, D. & Endres, R. G. Memory improves precision of cell sensing in fluctuating environments. *Sci. Rep.* **4,** 5688 (2014).
306. Vergassola, M., Villermaux, E. & Shraiman, B. I. 'Infotaxis' as a strategy for searching without gradients. *Nature* **445,** 406–9 (2007).
307. Govern, C. C. & Wolde, P. R. Ten. Optimal resource allocation in cellular sensing systems. *Proc. Natl. Acad. Sci.* **111,** 17486–17491 (2014).

# 2 Selected publications: Cell motility and motility control

## 2.1 "Flagellar synchronization independent of hydrodynamic interactions"

**Abstract.** Inspired by the coordinated beating of the flagellar pair of the green algae *Chlamydomonas*, we study theoretically a simple, mirror-symmetric swimmer, which propels itself at low Reynolds number by a revolving motion of a pair of spheres. We show that perfect synchronization between these two driven spheres can occur due to the motion of the swimmer and local hydrodynamic friction forces. Hydrodynamic interactions, though crucial for net propulsion, contribute little to synchronization for this free-moving swimmer.

## 2.2 "Cell body rocking is a dominant mechanism for flagellar synchronization in a swimming green alga"

**Abstract.** The unicellular green algae *Chlamydomonas* swims with two flagella, which can synchronize their beat. Synchronized beating is required to swim both fast and straight. A long-standing hypothesis proposes that synchronization of flagella results from hydrodynamic coupling, but the details are not understood. Here, we present realistic hydrodynamic computations and high-speed tracking experiments of swimming cells that show how a perturbation from the synchronized state causes rotational motion of the cell body. This rotation feeds back on the flagellar dynamics via hydrodynamic friction forces and rapidly restores the synchronized state in our theory. We calculate that this 'cell body rocking' provides the dominant contribution to synchronization in swimming cells, whereas direct hydrodynamic interactions between the flagella contribute negligibly. We experimentally confirmed the coupling between flagellar beating and cell body rocking predicted by our theory. We propose that the interplay of flagellar beating and hydrodynamic forces governs swimming and synchronization in *Chlamydomonas*.

## 2.3 "Active phase and amplitude fluctuations of the flagellar beat"

**Abstract.** The eukaryotic flagellum beats periodically, driven by the oscillatory dynamics of molecular motors, to propel cells and pump fluids. Small, but perceivable fluctuations in the beat of individual flagella have physiological implications for synchronization in collections of flagella as well as for hydrodynamic interactions between flagellated swimmers. Here, we characterize phase and amplitude fluctuations of flagellar bending waves using shape mode analysis and limit-cycle reconstruction. We report a quality factor of flagellar oscillations, $Q = 38.0 \pm 16.7$ (mean±s.e.). Our analysis shows that flagellar fluctuations are dominantly of active origin. Using a minimal model of collective motor oscillations, we demonstrate how the stochastic dynamics of individual motors can give rise to active small-number fluctuations in motor-cytoskeleton systems.

## 2.4 "Sperm navigation in 3D chemoattractant landscapes"

**Abstract.** Sperm require a sense of direction to locate the egg for fertilization. They follow gradients of chemical and physical cues provided by the egg or the oviduct. However, the principles underlying three-dimensional (3D) navigation in chemical landscapes are unknown. Here using holographic microscopy and optochemical techniques, we track sea urchin sperm navigating in 3D chemoattractant gradients. Sperm sense gradients on two timescales, which produces two different steering responses. A periodic component, resulting from the helical swimming, gradually aligns the helix towards the gradient. When incremental path corrections fail and sperm get off course, a sharp turning manoeuvre puts sperm back on track. Turning results from an 'off' Ca2+ response signifying a chemoattractant stimulation decrease and, thereby, a drop in cyclic GMP concentration and membrane voltage. These findings highlight the computational sophistication by which sperm sample gradients for deterministic klinotaxis. We provide a conceptual and technical framework for studying microswimmers in 3D chemical landscapes.

# 3 Selected publications: Self-organized pattern formation in cells and tissues

## 3.1 "Sarcomeric pattern formation by actin cluster coalescence"

**Abstract.** Contractile function of striated muscle cells depends crucially on the almost crystalline order of actin and myosin filaments in myofibrils, but the physical mechanisms that lead to myofibril assembly remains ill-defined. Passive diffusive sorting of actin filaments into sarcomeric order is kinetically impossible, suggesting a pivotal role of active processes in sarcomeric pattern formation. Using a one-dimensional computational model of an initially unstriated actin bundle, we show that actin filament treadmilling in the presence of processive plus-end crosslinking provides a simple and robust mechanism for the polarity sorting of actin filaments as well as for the correct localization of myosin filaments. We propose that the coalescence of crosslinked actin clusters could be key for sarcomeric pattern formation. In our simulations, sarcomere spacing is set by filament length prompting tight length control already at early stages of pattern formation. The proposed mechanism could be generic and apply both to premyofibrils and nascent myofibrils in developing muscle cells as well as possibly to striated stress-fibers in non-muscle cells.

## 3.2 "Scaling and regeneration of self-organized patterns"

**Abstract.** Biological patterns generated during development and regeneration often scale with organism size. Some organisms e.g. flatworms can regenerate a re-scaled body plan from tissue fragments of varying sizes. Inspired by these examples, we introduce a generalization of Turing patterns that is self-organized and self-scaling. A feedback loop involving diffusing expander molecules regulates the reaction rates of a Turing system, thereby adjusting pattern length scales proportional to system size. Our model captures essential features of body plan regeneration in flatworms as observed in experiments.

# 4 Contribution of the author in collaborative publications

The author is principal author of all publications presented in this thesis. He conceived the original idea for the projects in all-but-one cases. The only exception is publication 2.4: "Sperm navigation …", which represent a combination of experiment and theory on sperm chemotaxis, where the experimental part was initiated first. The theoretical part of this publication has been solely contributed by the author. Below, we list the specific contributions of the author for all publications presented in this thesis.

2.1: "Flagellar synchronization independent of hydrodynamic interactions"
- **contribution of author:** original conception of project, development of theoretical description, all analytic calculations, all stochastic simulations, manuscript preparation including all figures

2.2: "Cell body rocking is a dominant mechanism for flagellar synchronization in a swimming alga"
- **contribution of author:** original conception of project, management of theory-experiment collaboration, development of theoretical description, all analytic calculations, all hydrodynamic computations, development of image analysis software, data analysis, manuscript preparation including all figures

2.3: "Active phase and amplitude fluctuations of the flagellar beat"
- **contribution of author:** original conception of project, development of data analysis method for limit cycle reconstruction, data analysis, development of theoretical description, development of analytical theory together with Rui Ma, manuscript preparation including all figures

2.4: "Sperm navigation in 3D chemical landscapes"
- **contribution of author:** development of theoretical description, hydrodynamic computations of chiral flagellar swimming, algorithm development for data analysis (track smoothing, helix fitting, chemoattractant diffusion, time series analysis), comparison of theory and experiment, writing of theoretical part of manuscript, contribution to introduction and discussion, prepared figure panels (Fig. 1c, Fig. 2, Fig. 3c, Fig. 4c, Fig. 4f, Fig. 6, Fig. S3b Fig. S3f, Fig. S4)

3.1: "Sarcomeric pattern formation by actin cluster coalescence"
- **contribution of author:** original conception of project, development of theoretical description, all stochastic simulations, all analytic calculations, manuscript preparation including all figures

3.2: "Scaling and regeneration of self-organized patterns"
- **contribution of author:** original conception of project, development of initial theoretical description, close supervision of PhD student Steffen Werner who formulated the final description, coordination of the project, manuscript preparation and conception of all figures

# 5 Eidesstattliche Versicherung

Hiermit versichere ich an Eides statt, dass ich die vorliegende Arbeit ohne unzulässige Hilfe Dritter und ohne Benutzung anderer als der angegebenen Hilfsmittel angefertigt habe. Die aus fremden Quellen direkt oder indirekt übernommenen Gedanken sind als solche kenntlich gemacht.

Die der Habilitationsschrift zugrunde liegenden Publikationen wurden in Zusammenarbeit mit den genannten Autoren angefertigt. Dabei ging jeweils die initiale Idee auf mich zurück. Die Projekte wurden durch mich koordiniert und die Publikationen von mir geschrieben. Die einzige Ausnahme bildet Publikation 2.4 zur Spermien-Chemotaxis, welche Experiment und Theorie kombiniert. Hier beschränkt sich mein Beitrag auf den Theorieteil, d.h. die Entwicklung der theoretischen Beschreibung, alle numerischen Simulationen, sowie die quantitative Auswertung der experimentellen Daten. Eine detaillierte Auflistung meines Beitrags zu den Publikationen findet sich auf Seite 63 dieser Habilitationsschrift.

Diese Arbeit wurde bisher weder im Inland noch im Ausland in gleicher oder ähnlicher Form einer anderen Prüfungsbehörde vorgelegt. Ich habe bisher kein Habilitationsgesuch an anderen Hochschulen eingereicht. Ein Führungszeugnis gemäß § 30 Abs. 5 Bundeszentralregistergesetz wurde bei der zuständigen Meldebehörde beantragt und wird an die Fakultät übersendet.

Dresden, den 23.05.2016

Benjamin M. Friedrich

# 6  Appendix: Reprints of publications

**Publication 2.1:** "Flagellar synchronization independent of hydrodynamic interactions"
Published in:
*Phys. Rev. Lett.* **109**(13), p. 138102, 2012;
5 pages, 4 figures.

**Publication 2.2:** "Cell body rocking is a dominant mechanism for flagellar synchronization in a swimming green alga"
Published in:
*Proc. Natl. Acad. Sci. U.S.A.* **110**(45), 18058(6). 2013;
6 pages, 5 figures, 12 pages of supporting material.

**Publication 2.3:** "Active phase and amplitude fluctuations of the flagellar beat"
Published in:
*Phys. Rev. Lett.* **113**(4), 048101, 2014;
5 pages, 3 figures, 5 pages of supporting material.

**Publication 2.4:** "Sperm navigation in 3D chemoattractant landscapes"
Published in:
*Nature Communications* **6**, 7985, 2015;
10 pages, 6 figures, 14 pages of supporting material.

**Publication 3.1:** "Sarcomeric pattern formation by actin cluster coalescence"
Published in:
*PLoS Comp. Biol.* **8**(6), e1002544, 2012;
10 pages, 6 figures, 11 pages of supporting material.

**Publication 3.2:** "Scaling and regeneration of self-organized patterns"
Published in:
*Phys. Rev. Lett.* **114**(13), 138101, 2015;
5 pages, 3 figures.